\documentclass[useAMS,usenatbib]{mn2e}

\usepackage{graphicx}
\usepackage{graphics}
\usepackage{color}
\usepackage{url}
\usepackage[caption=false]{subfig}
\usepackage{verbatim}
\usepackage{array}
\usepackage{amssymb}
\usepackage{amsmath}
\usepackage{longtable}
\usepackage{pbox}
\usepackage{ulem}

\newcommand{\name}{ASASSN-15lh}
\newcommand{\swift}{\it Swift}

\newcommand{\editfinal}[1]{\textcolor{black}{#1}}
\newcommand{\revone}[1]{\textcolor{black}{#1}}

\newcommand{\ion}[2]{{#1~\uppercase \expandafter{\romannumeral #2}}}
\newcolumntype{C}[1]{>{\centering\arraybackslash}m{#1}}
\newcolumntype{L}[1]{>{\arraybackslash}m{#1}}

\begin{document}

\title[]{The Unexpected, Long-Lasting, UV Rebrightening of the Super-Luminous Supernova {\name}}

\author[D.~Godoy-Rivera et al.]{D.~Godoy-Rivera$^{1}$\thanks{e--mail:godoyrivera.1@osu.edu}, K.~Z.~Stanek$^{1,2}$, C.~S.~Kochanek$^{1,2}$, Ping Chen$^{3}$, Subo~Dong$^{4}$,
\newauthor
J.~L.~Prieto$^{5,6}$, B.~J.~Shappee$^{7,8}$, S.~W.~Jha$^{9}$, R.~J.~Foley$^{10}$, Y.-C.~Pan$^{11}$,
\newauthor
T.~W.-S.~Holoien$^{1,2}$, Todd.~A.~Thompson$^{1,2}$, D.~Grupe$^{12}$ and J.~F.~Beacom$^{1,2,13}$ \\
  $^{1}$  Department of Astronomy, The Ohio State University, 140 West 18th Avenue, Columbus, OH 43210, USA \\
  $^{2}$  Center for Cosmology and AstroParticle Physics (CCAPP), The Ohio State University, 191 W. Woodruff Ave., Columbus, OH 43210, USA \\
  $^{3}$  Department of Astronomy, Peking University, Yi He Yuan Road 5, Hai Dian District, Beijing 100871, China\\
  $^{4}$  Kavli Institute for Astronomy and Astrophysics, Peking University, Yi He Yuan Road 5, Hai Dan District, Beijing, China \\
  $^{5}$  N\'ucleo de Astronom\'ia de la Facultad de Ingenier\'ia y Ciencias, Universidad Diego Portales, Av. Ej\'ercito 441, Santiago, Chile \\
  $^{6}$  Millennium Institute of Astrophysics, Santiago, Chile \\
  $^{7}$  Carnegie Observatories, 813 Santa Barbara Street, Pasadena, CA 91101, USA \\
  $^{8}$  Hubble and Carnegie-Princeton Fellow\\
  $^{9}$  Department of Physics and Astronomy, Rutgers, The State University of New Jersey, 136 Frelinghuysen Road, Piscataway, NJ 08854, USA \\
  $^{10}$ Department of Astronomy and Astrophysics, University of California, Santa Cruz, CA 95064, USA\\
  $^{11}$ Astronomy Department, University of Illinois at Urbana-Champaign, 1002 W. Green Street, Urbana, IL 61801, USA \\
  $^{12}$ Department of Earth and Space Science, Morehead State University, 235 Martindale Dr., Morehead, KY 40351, USA \\
  $^{13}$ Department of Physics, The Ohio State University, 191 W. Woodruff Ave, Columbus, OH 43210, USA \\
}

\maketitle
\begin{abstract}
\revone{Given its peak luminosity and early-time spectra, {\name} was classified as the most luminous supernova (SN) ever discovered \citep{dong16}}. Here we report a UV rebrightening of {\name} observed with {\swift} during our follow-up campaign. The rebrightening began at $t \simeq$ 90 days (observer frame) after the primary peak and was followed by a $\sim 120$-day long plateau in the bolometric luminosity, before starting to fade again at $t\simeq 210$ days. \revone{{\name}} rebrightened in the {\swift} UV bands by $\Delta m_{UVW2} \simeq -1.75$ mag, $\Delta m_{UVM2} \simeq -1.25$ mag, and $\Delta m_{UVW1} \simeq -0.8$ mag, but did not rebrighten in the optical bands. Throughout its initial decline, subsequent rebrightening, and renewed decline, the spectra did not show evidence of interactions between the ejecta and circumstellar medium (CSM) such as narrow emission lines. There are hints of weak H$\alpha$ emission at late-times, \revone{but \citet{margutti16} have shown that it is narrow line emission consistent with star formation in the host nucleus}. By fitting a blackbody we find that during the rebrightening the effective photospheric temperature increased from $T_{BB} \simeq 11000$ K to $T_{BB} \simeq 18000$ K. Over the $\sim$ \editfinal{\revone{550}} days since its detection, {\name} has radiated $\sim \editfinal{\revone{1.7 \textendash 1.9}} \times 10^{52}$ ergs. \revone{Although its physical nature remains uncertain, the evolution of {\name}'s photospheric radius, its radiated energy, and the implied event rate, are all more similar to those of H-poor superluminous supernovae (SLSNe-I) than to tidal disruption events (TDEs). }
\end{abstract}


\section{Introduction}
\label{sec:intro}

The discovery of super-luminous supernovae (SLSNe) \citep{quimby11,gal-yam12} came as a surprise to the astronomical community despite many decades of research on supernovae (SNe). One reason is that SLSNe appear predominantly in dwarf galaxies \citep{neill11,stoll11,lunnan15}, which were not well observed in early, local, targeted SN surveys that concentrated on large star-forming galaxies. SLSNe are now being systematically observed by wide-field surveys, starting with the Texas Supernova Search (TSS; \citealt{quimby06}), and now including the All-Sky Automated Survey for SuperNovae (ASAS-SN\footnote{\url{http://www.astronomy.ohio-state.edu/~assassin/}}; \citealt{shappee14}), the Palomar Transient Factory (PTF; \citealt{law09,rau09}), and the Panoramic Survey Telescope and Rapid Response System (Pan-STARRS; \citealt{kaiser02}). SLSNe, while rare, have the potential to illuminate our understanding of all terminal stellar explosions, not just the most extreme. 

As the community was getting used to the extreme being normal with SLSNe, we discovered {\name} \citep{dong16}, which strains theories for SLSN explosion mechanisms. At a redshift of $z=0.2326$, {\name} was discovered on 2015 June 14 by ASAS-SN, a project designed to monitor the entire visible sky on rapid cadence, looking for bright transients \citep{shappee14}. {\name} reached a peak bolometric luminosity of $(2.2 \pm 0.2) \times 10^{45}$ ergs s$^{-1}$ and radiated $(1.1 \pm 0.2) \times 10^{52}$ ergs in the first four months of observations \citep{dong16}. Its \revone{early-time} spectra are reasonably matched by the spectra of type I (H-poor) SLSNe (SLSNe-I). Its host, APMUKS(BJ) B215839.70−615403.9 \citep{maddox90}, in contrast to the typical hosts of SLSNe, is a luminous ($M_K\simeq -25.5$) galaxy with a low star-formation rate and a high total mass. 

In this paper we report a significant \revone{UV} rebrightening episode, initially noted by \citet{brown15atel}, lasting $\sim 160$ days \revone{and dominated by} a $\sim$120-day long plateau in the bolometric luminosity. This rebrightening, although unexpected, should help to constrain models for the most luminous SN ever discovered, and provide new insights into the poorly understood physics and mechanisms that power SLSNe-I.

In  \revone{\S}\ref{sec:obs} we present photometric and spectroscopic data taken during our follow-up campaign. In \revone{\S}\ref{sec:SED_LTR} we use these data to model the temporal evolution of the bolometric luminosity, effective photospheric temperature and radius of the source, as well as its spectral energy distribution (SED). In \revone{\S}\ref{sec:discussion} we calculate the energy radiated by {\name}, \revone{briefly comment on the still unknown power source and nature of the rebrightening, and compare its properties to the tidal disruption events (TDEs) discovered by ASAS-SN and other SLSNe. We summarize our results in \S\ref{sec:summary}.}

\begin{figure*}
\begin{minipage}{\textwidth}
\centering
\subfloat{{\includegraphics[width=1.0\linewidth]{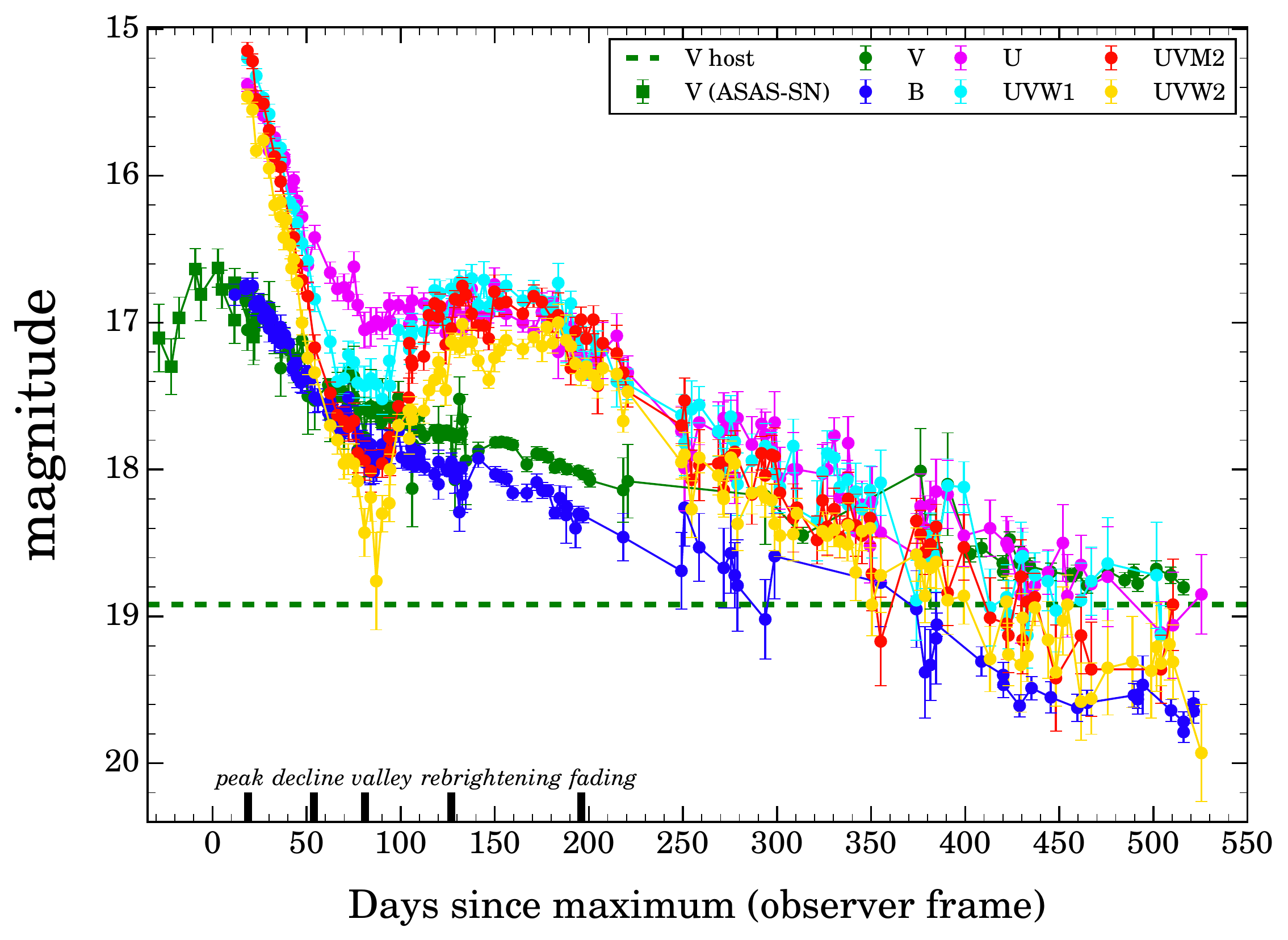}}}
\caption{Light curves of {\name} in the observer frame spanning from 29 days before maximum through \editfinal{\revone{525}} days after maximum (MJD 57150 to MJD \editfinal{\revone{57704}}). The {\swift} and LCOGT follow-up data are shown as filled circles, and the ASAS-SN $V$-band data are shown as green squares. The synthetic host galaxy magnitude of $V=18.92$ \citep{dong16} is shown as the green horizontal dashed line. All magnitudes are in the Vega system as measured by {\swift}, and were not corrected for extinction or the host contribution. Representative dates for the different stages of the light curve evolution (see \revone{\S}\ref{sec:photometric_obs}) are shown as black vertical lines at the bottom.}
\label{fig:fig1}
\end{minipage}
\end{figure*}


\section{Photometry and Spectroscopy}
\label{sec:obs}

In this section we present our follow-up photometric and spectroscopic observations of {\name}. For consistency we describe the complete period of observation ($\sim$ \editfinal{\revone{550}} days) \revone{from} the first detection of the transient \revone{on} MJD 57150 through \editfinal{\revone{MJD 57704}}. The data for the period before MJD 57283.4 were already presented in \citet{dong16}. Between MJD 57399.5 and MJD 57427.5 our follow-up campaign was \revone{suspended} due to Sun constraints.

\subsection{Photometric Observations}
\label{sec:photometric_obs}

{\name} was extensively observed by {\swift}. We use the series of X-ray telescope (XRT; \citealt{burrows05}) and UV/Optical telescope (UVOT; \citealt{roming05}) target-of-opportunity (ToO) observations taken between MJD 57197 and MJD \editfinal{\revone{57704}}. The UVOT observations were obtained in six filters: $V$ (5468 \AA), $B$ (4392 \AA), $U$ (3465 \AA), $UVW1$ (2600 \AA), $UVM2$ (2246 \AA), and $UVW2$ (1928 \AA) {\citep{poole08}. To extract the source and sky counts we used  $\sim$ 5\farcs0 and $\sim$ 40\farcs0 radii regions, respectively, using the software task {\sc uvotsource}. The measured count rates were converted into magnitudes and fluxes using standard UVOT calibrations \citep{poole08,breeveld10}. The magnitudes obtained here, especially \revone{for} the UV bands, are slightly different from the ones in \citet{dong16} because of updates to the {\sc uvotsource} calibration database. The {\swift} photometric data are presented in Table \ref{tab:table1}.

During the observations, the XRT operated in Photon Counting mode \citep{hill04}. The X-ray data taken between MJD 57197 and MJD \editfinal{\revone{57704}} were combined and reduced using the software tasks {\sc xrtpipeline} and {\sc xselect} to produce a 0.3 \textendash 10  keV image. The total exposure time is \editfinal{\revone{232}} ks. To extract the source and sky counts we used regions of radius 10 pixels ($\sim$ 23\farcs6) and 100 pixels ($\sim$ 236$\arcsec$), respectively. We do not detect X-ray emission from {\name} with a $3\sigma$ upper limit of $\sim \editfinal{\revone{1.1}} \times 10^{-4}$ counts s$^{-1}$ using the Bayesian inference method described by \citet{kraft91}. Assuming a standard power law spectrum with $\Gamma=2$, this implies a flux limit of $< \editfinal{\revone{6}} \times 10^{-15}$ ergs cm$^{-2}$ s$^{-1}$, or given the host luminosity distance of $d \simeq 1170$ Mpc \citep{dong16}, a luminosity limit of $L_{X} \leq \editfinal{\revone{9.8 \times10^{41}}} $ ergs s$^{-1}$ ($\leq \editfinal{\revone{2.6 \times 10^{8}}}$ L$_{\odot}$) \revone{(see also \citealt{brown16asassn15lh})}.

Only the combined X-ray image was considered since the constraints for individual epochs are weaker. In \citet{dong16} we used the same procedure to derive a $3\sigma$ X-ray flux limit of $<1.6 \times 10^{-14}$ ergs cm$^{-2}$ s$^{-1}$ in the same energy range from a total exposure of 81 ks between MJD 57197 and MJD 57283. The updated X-ray limit presented here is more stringent than the one reported in \citet{dong16} due to the longer observation period. \revone{This limit is consistent with the X-ray detection of {\name} at a flux of $\sim 4 \times 10^{-15}$ ergs cm$^{-2}$ s$^{-1}$ reported by \citet{margutti16}. The origin of the X-ray emission is unclear, and time variability is likely required for a useful interpretation.}

In addition to the {\swift} observations, we also obtained $V$- and $B$-band photometry with the Las Cumbres Observatory Global Telescope Network (LCOGT; \citealt{brown13}) 1-m telescopes located at Siding Spring Observatory, South African Astronomical Observatory, and Cerro Tololo Inter-American Observatories. The images were reduced following standard procedures. We measured instrumental magnitudes using aperture photometry with a 6-pixel aperture radius ($\sim$ 3\farcs0) and then used AAVSO Photometric All-Sky Survey (APASS; \citealt{henden15}) reference stars in the field to obtain calibrated Johnson-Cousins magnitudes for our target. The resulting magnitudes are slightly different from the ones of \citet{dong16} because here we used a larger number of reference stars in the calibration. The LCOGT photometric data are presented in Table \ref{tab:table2}.

As explained in \citet{dong16}, we also have ASAS-SN $V$-band pre-discovery images that allow us to date the detection of {\name} to MJD 57150. These data provide the only photometric observations for the period before MJD 57189.4.

Given that the $V$- and $B$-band of {\swift} and LCOGT are not identical, we translated the LCOGT magnitudes to the {\swift} system by deriving offsets between the two data sets. The offsets were calculated using linear interpolation of the LCOGT measurements for the initial period where we have overlapping observations (MJD 57196.5 to MJD 57308.5). We obtained $\Delta V=-0.052$ and $\Delta B=+0.026$, where $\Delta m = m_{{\it Swift}} - m_{\text{LCOGT}}$.

Figure \ref{fig:fig1} shows the light curves of {\name} from its first detection at MJD 57150 through MJD \editfinal{\revone{57704}}. The source magnitudes shown in Figure \ref{fig:fig1} are not corrected for the host contribution or Galactic extinction. All $V$- and $B$- band magnitudes are shown in the {\swift} system. The synthetic $V$-band magnitude of the host galaxy that we derived in \citet{dong16} by modeling its SED is also shown. Hereafter the times will be expressed relative to the maximum of the ASAS-SN $V$-band light curve (MJD 57178.5; \citealt{dong16}) in the observer frame, unless otherwise stated.

The evolution of the light curve is striking. After the maximum, {\name} showed a continuous decline at all wavelengths. Then, near $t \simeq$ 90 days, a rebrightening began in the UV bands. Figure \ref{fig:fig1} shows that the rebrightening is largest for the bluer bands: $\Delta m_{UVW2} \simeq -1.75$ mag (peak at $t\simeq 133$ days), $\Delta m_{UVM2} \simeq -1.25$ mag (peak at $t\simeq 133$ days), and $\Delta m_{UVW1} \simeq -0.8$ mag (peak at $t\simeq 138$ days). While this happened in the UV, the $U$-band remained roughly constant for 97 days (a standard deviation of $\sigma_{U} \simeq 0.08$ mag between $t\simeq 84$ days and $t\simeq 181$ days), and the $V$- and $B$-band continued to slowly decline. 

We identify five distinct phases in the multi-wavelength light curve of {\name}, and note corresponding representative dates: the {\it peak} ($t\simeq 19$ days), where all the bands had their maximum brightness, except for the $V$-band (which had its maximum around $t=0$ in the ASAS-SN data); a following {\it decline} ($t\simeq 54$ days), where all the bands declined roughly monotonically; the {\it valley} ($t\simeq 81$ days), where the UV+$U$ bands reached their first minimum; the unexpected {\it rebrightening} ($t\simeq 127$ days) that was seen in the UV and not in the optical (although the $U$-band stopped declining); and finally a {\it fading} stage ($t\simeq 196$ days), where all bands again declined monotonically. The representative dates of these stages are shown as black vertical lines at the bottom of Figure \ref{fig:fig1}. In \revone{\S}\ref{sec:spectroscopic_obs} we show spectra selected to approximately match these representative dates.

\subsection{Spectroscopic Observations}
\label{sec:spectroscopic_obs}

\begin{figure}
\centering
\subfloat{{\includegraphics[width=1.0\linewidth]{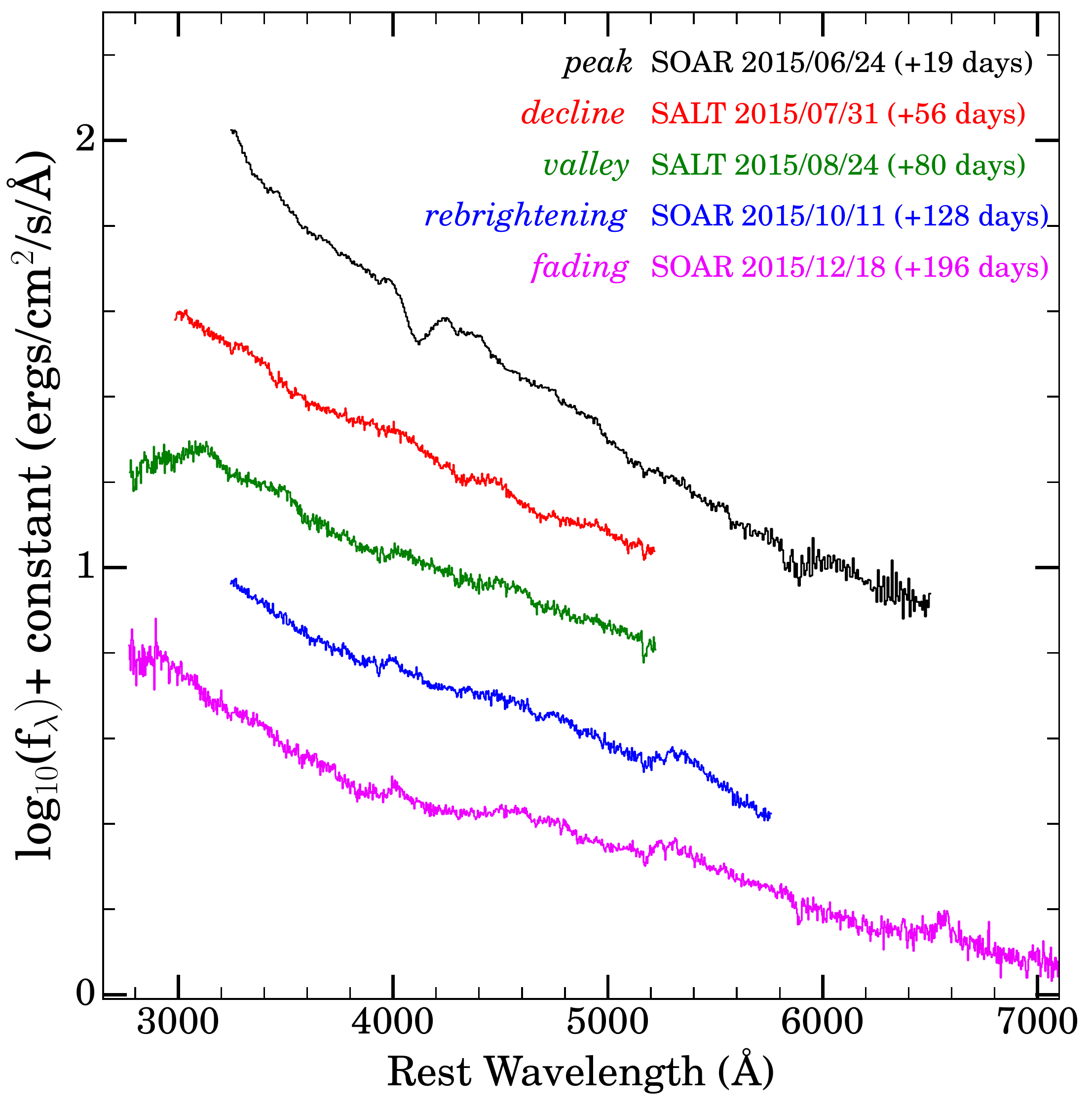}}}
\caption{Rest frame spectroscopic evolution of {\name}. Spectra dates were selected to represent the different stages of the light curve evolution of Figure \ref{fig:fig1} (see \revone{\S} \ref{sec:photometric_obs}). The dates of the observations in UT time, and with respect to the light curve maximum (in the observer frame), are shown together with the instrument used and the name of the phase. The colors of the spectra match the corresponding date and instrument. Offsets have been added to the spectra for clarity.}
\label{fig:fig2}
\end{figure}

\begin{figure}
\centering
\subfloat{{\includegraphics[width=1.0\linewidth]{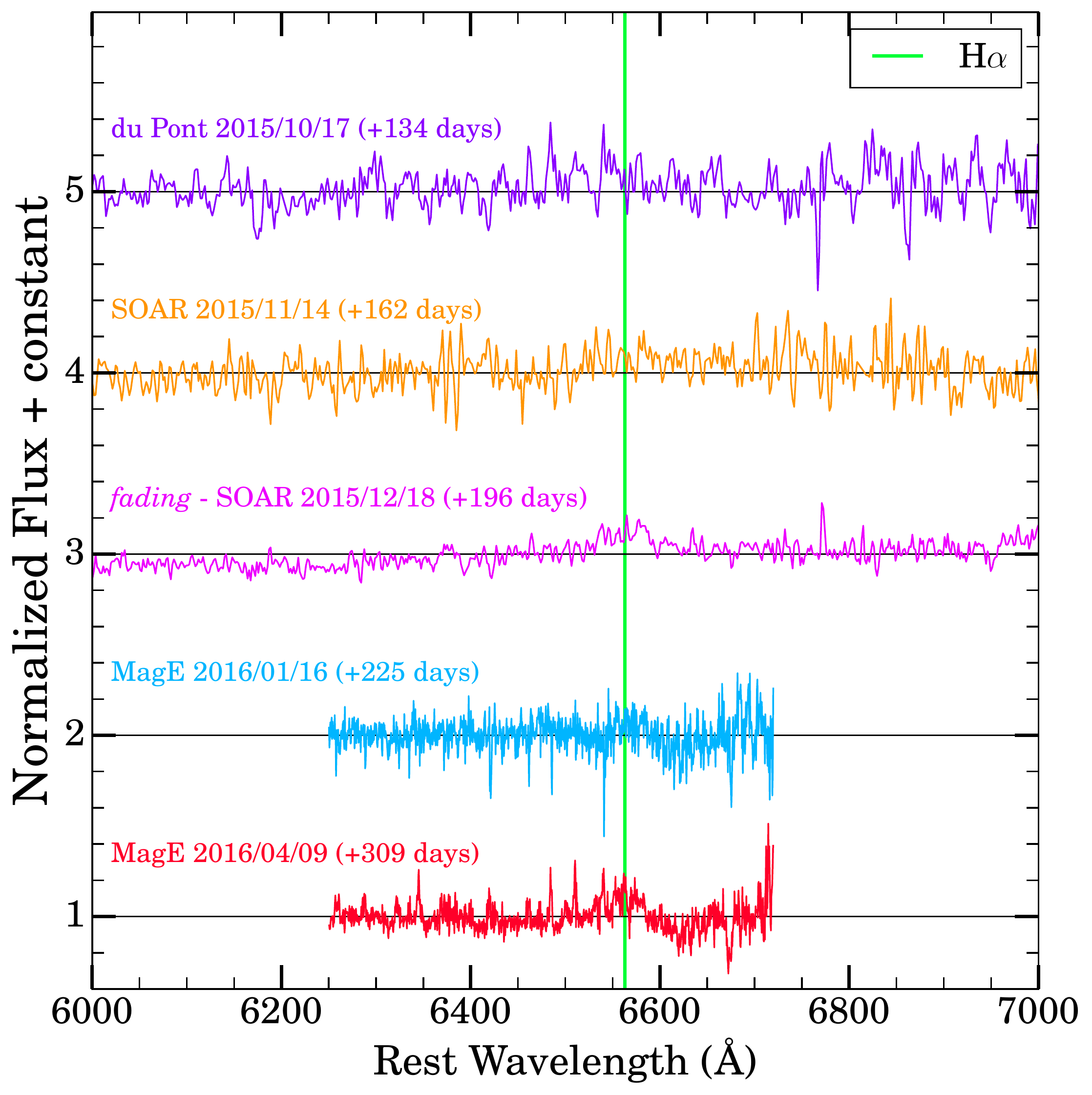}}}
\caption{Rest frame spectroscopic evolution of {\name} near \revone{the wavelength of }H$\alpha$. The spectra were normalized by a 2nd-degree polynomial fit to the continuum to emphasize on the relative strength of any potential features. The spectra were selected to cover the period before and after the SOAR spectrum of the {\it fading} stage ($t=196$ days). The dates of the observations in UT time, and with respect to the light curve maximum (in the observer frame), are shown together with the instrument used. For the MagE spectra, only the orders containing the wavelength of H$\alpha$ were used in the continuum fit. The colors of the spectra match the corresponding date and instrument. \revone{The spectra have an arbitrary flux offset and a horizontal line is drawn at the mean of each spectrum}. The expected position of H$\alpha$ (6562.8 \AA) is shown as the vertical green line.}
\label{fig:fig3}
\end{figure}

In the \revone{first} $\sim$ \editfinal{330} days after the detection of {\name}, we obtained numerous spectra. Here we show only a representative sample, illustrating the five stages defined in \revone{\S}\ref{sec:photometric_obs}, and other late-time epochs. Spectra were obtained with the Robert Stobie Spectrograph \citep{smith06} on the 10-m Southern African Large Telescope (SALT), the Goodman Spectrograph \citep{clemens04} on the 4-m Southern Astrophysical Research Telescope (SOAR), the Magellan Echellette (MagE; \citealt{marshall08}) Spectrograph on the 6.5-m Baade telescope, and the Wide Field CCD (WFCCD) camera on the 2.5-m du Pont telescope.

The SALT data were taken with a 1\farcs5 wide longslit (oriented to PA$=$127.7$^\circ$ east of north) and used the PG0900 grating, resulting in a spectral resolution of $\lambda/\Delta\lambda \approx 900$. The 2015 July 31 and 2015 August 24 observations used two tilt positions to cover the wavelength range from 3500 \textendash 6400 \AA\ (without chip gaps) with a total exposure time of 2240 s. The data were reduced with a custom pipeline that incorporates routines from PyRAF and PySALT \citep{crawford10}. 

The SOAR observations used an 1\farcs07 wide longslit with the ``M1'' blue (for all epochs) and ``M2'' red (for the December epoch; the GG455 order-blocking filter was also used) 400-line VPH gratings. Standard CCD processing and spectral extraction were accomplished with IRAF\footnote{IRAF: the Image Reduction and Analysis Facility is distributed by the National Optical Astronomy Observatory, which is operated by the Association of Universities for Research in Astronomy, Inc.\ (AURA) under cooperative agreement with the National Science Foundation (NSF).}.  The data were extracted using the optimal algorithm of \citet{horne86}.  Low-order polynomial fits to calibration-lamp spectra were used to establish the wavelength scale, and small adjustments derived from night-sky lines in the object frames were applied.  We employed our own IDL routines to flux calibrate the data and remove telluric lines using the well-exposed continua of spectrophotometric standards \citep{wade88, foley03}.  Details of our spectroscopic reduction techniques are described by \citet{silverman12}.

We obtained two medium-resolution (R $\simeq$ 4000) MagE optical spectra of {\name} on 2016 Jan 16 and 2016 April 9. We extracted the echellette orders using Dan Kelson's MagE pipeline\footnote{\url{http://code.obs.carnegiescience.edu/mage-pipeline}}. The 2015 October 17 du Pont spectrum was reduced using standard IRAF procedures.

Figure \ref{fig:fig2} shows the spectroscopic evolution of {\name}. The shape of the spectra indicate that the source remained hot during the whole period of observation, in agreement with the blackbody fit ($T_{\text{BB}} \gtrsim$ 11000 K) to the photometric data presented in \revone{\S}\ref{sec:SED_LTR}. 

For the first $\sim$ 45 days the most prominent feature was the absorption line near $\sim$ 4100 \AA\ (see \revone{\S\ref{sec:discussion3_spec} and} also Figure 2 of \citealt{dong16}). \revone{This feature has also been seen in the SLSNe-I SN 2010gx \citep{pastorello10}, SN 2015bn \citep{nicholl16a}, and PTF12dam \citep{nicholl13} among others, and it has been attributed to \ion{O}{2} \citep{quimby11,inserra13}.} The feature almost completely disappeared $\sim$ 35 days before the rebrightening began. 

From the {\it decline} stage \revone{onward}, the spectra show a mostly featureless continuum at all epochs, without the appearance of significant features, with the exception of the last spectrum. This SOAR spectrum taken at $t = 196$ days (corresponding to the {\it fading} stage) shows a slight bump near the expected position of H$\alpha$ (6562.8 \AA). No such feature is seen at early-times (see Figure 2 of \citealt{dong16}). 

Since the absence of H and He emission features is one of the key points for the SLSN-I hypothesis, this feature requires further investigation. Figure \ref{fig:fig3} shows a selection of spectra chosen to display the late-time evolution of {\name} near the wavelength of H$\alpha$. The spectra were normalized by a smooth continuum to study the relative strength of any feature. The SOAR spectrum at $t=196$ days is the only one with significant evidence for a bump near H$\alpha$. The du Pont spectrum at $t=134$ days, and the SOAR spectrum at $t=162$ days, do not show significant features at that wavelength. The MagE spectra at $t=225$ days and at $t=309$ days show hints of features near the H$\alpha$ wavelength, but are also strongly affected by the quality of the telluric corrections and noise. We cannot definitively determine if the feature seen near the wavelength of H$\alpha$ is an instrumental artifact, the effect of telluric lines, or a weak, transient broad H$\alpha$ emission line. 

\revone{\citet{milisavljevic15atel} also report no sign of broad H$\alpha$ emission at $t=135$ days, while \citet{leloudas16} claim a H$\alpha$ detection (FWHM $\simeq$ 2500 km s$^{-1}$) at $t=37$ days. If real, the broad H$\alpha$ emission would still be energetically negligible. It would be about $\sim 20$ times weaker relative to the continuum than the H$\alpha$ emission observed in the SLSN-I iPTF13ehe \citep{yan15}. More importantly, \citet{margutti16} clearly identify narrow H$\alpha$ and [\ion{N}{2}] emission in a spectrum taken at $t=301$ days, and convincingly argue that all the evidence for H$\alpha$ emission from the transient is simply a blend of these emission lines produced by star formation in the nucleus of the host galaxy.}


\section{SED, Luminosity and Effective Temperature Evolution}
\label{sec:SED_LTR}

As seen in Figure \ref{fig:fig1}, our UV monitoring was continuous except for the Sun constraint break ($221 \text{ days} < t< 249 \text{ days}$). \revone{In addition,} the {\swift} $V$ and $B$-band observations were interrupted between $t=135$ days and $t=188$ days due to \revone{their absence} in the ToO requests during that period, but the gap was covered by the LCOGT photometry. After $t=$188 days {\swift} began to observe the source in the optical bands again, although not always including both \revone{the} $V$ and $B$-bands \revone{simultaneously}. The LCOGT photometry was again used when needed.

We obtained fluxes in each band by correcting for Galactic extinction, assuming $E(B-V)=0.03$ mag \citep{schlafly11}, and subtracting the host galaxy contribution. The fluxes were fit to blackbody models using Markov Chain Monte Carlo (MCMC) methods to derive luminosities, photospheric effective temperatures, and apparent photospheric radii. While the light curves of Figure 1 include all the photometric data, for $t<221$ days (beginning of Sun constraint break) the SED calculations and blackbody fits only use epochs for which the {\swift} magnitudes satisfy $\sigma_m \leq 0.2$, the LCOGT magnitudes satisfy $\sigma_m \leq 0.10$, and the four UV bands have measured magnitudes. These quality cuts were not used for the period after the Sun constraint break, when the \revone{target} was fainter, given the lower signal-to-noise ratio of the observations. 

As in \citet{dong16}, for epochs earlier than $t=$ 12 days, when the photometry relied only on ASAS-SN $V$-band data, we used two different priors on the blackbody temperature. One extrapolates between $t=12.5$ days and $t=62.5$ days and increases linearly in logarithmic scale towards earlier times, while the other assumes a constant temperature (the one at $t=12.5$ days) for those epochs. After $t=12.5$ days we did not use any prior on the temperature because {\swift} started observing {\name} and UV photometry became available. The numerical values of the blackbody models are presented in Tables \ref{tab:table3}, \ref{tab:table4}, and \ref{tab:table5}. \revone{Our results for the bolometric luminosity roughly agree with those presented by \citet{brown16asassn15lh}}.

\begin{figure}
\centering
\subfloat{{\includegraphics[width=1.0\linewidth]{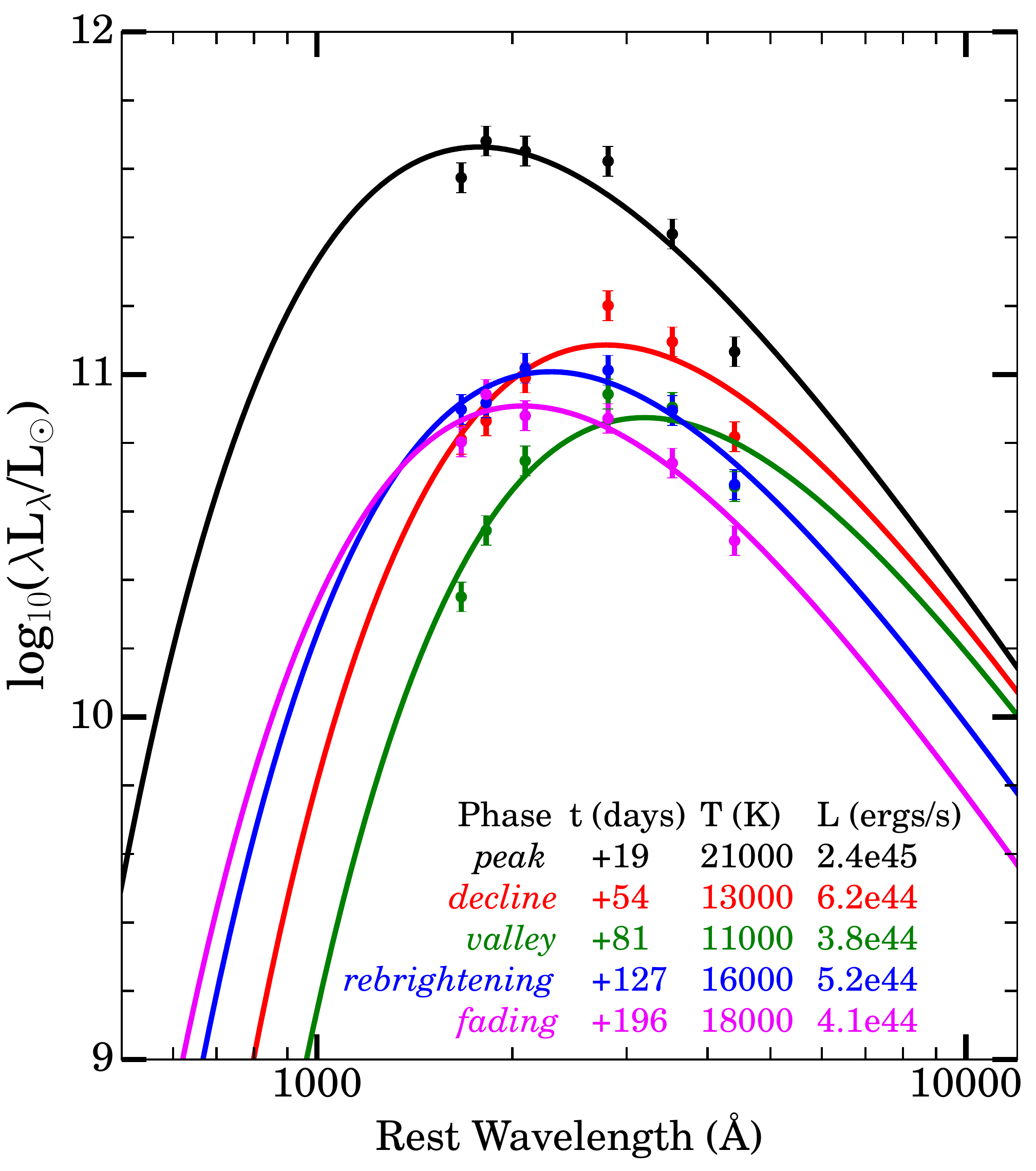}}}
\caption{SED evolution of {\name}. The observed SEDs are shown as filled circles while the best-fit blackbody models are shown as solid lines. 
These epochs correspond to the dates highlighted in Figure \ref{fig:fig1} (see \revone{\S}\ref{sec:photometric_obs}), and are within 2 days of the spectroscopy epochs shown in Figure \ref{fig:fig2}. \revone{The} names of the phases, dates (in the observer frame), best-fit effective temperatures, and luminosities, are listed for each epoch in the corresponding color. The X-ray luminosity upper limit derived in \revone{\S}\ref{sec:photometric_obs}, $\log_{10}(L_{\text{X}}/L_{\odot}) \leq \revone{8.4}$, and the radio luminosity upper limit reported by \citet{kool15atel}, $\log_{10}(L_{\text{Radio}}/L_{\odot}) \leq 4.9$, are below the scale of the figure.}
\label{fig:fig4}
\end{figure}

\begin{figure}
\centering
\subfloat{{\includegraphics[width=1.0\linewidth]{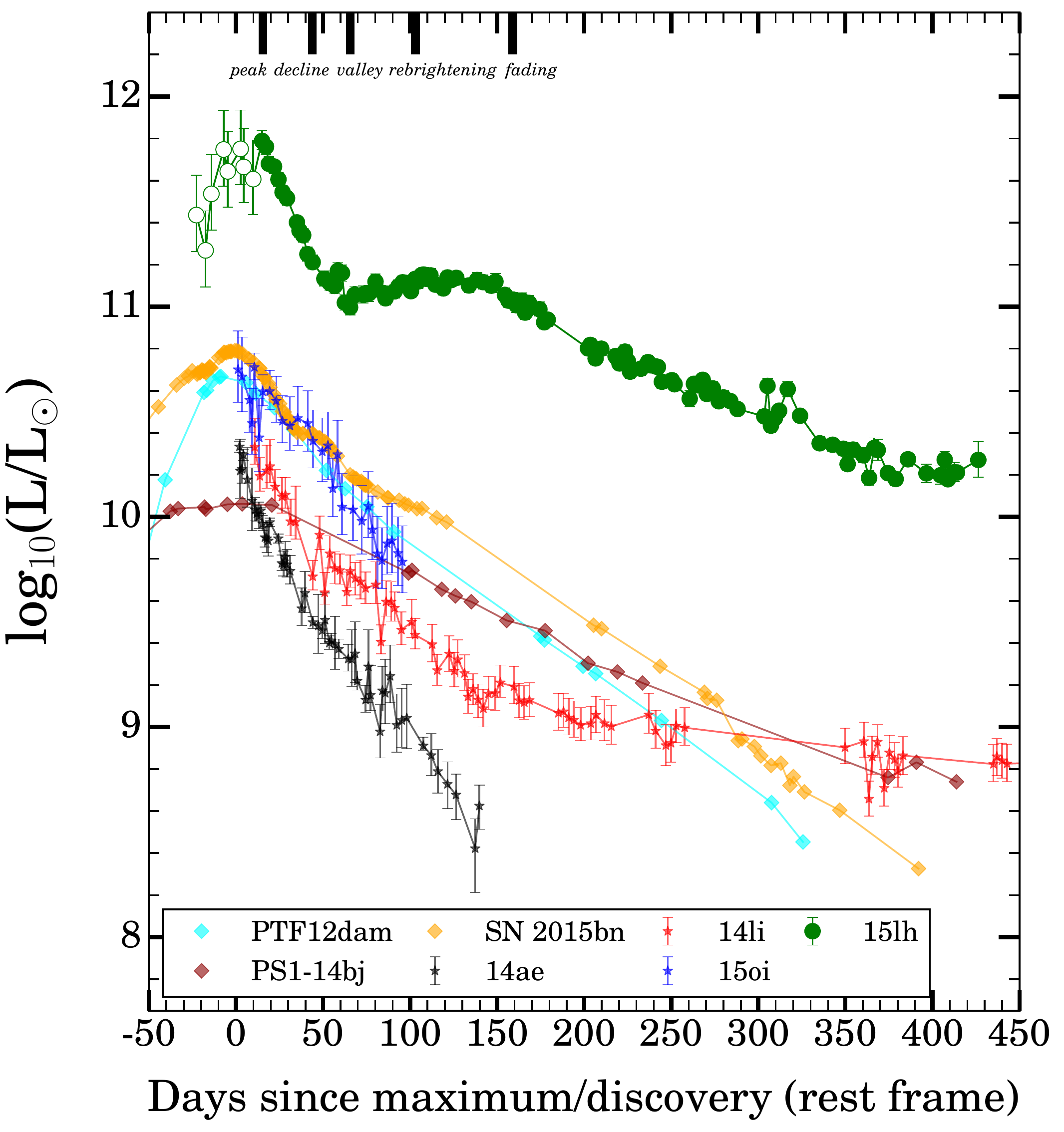}}}
\caption{Bolometric luminosity evolution of {\name} for blackbody fits using the constant-temperature prior (see \revone{\S}\ref{sec:SED_LTR}) at early-times (open green circles), and no prior afterwards (filled green circles). Representative dates for the different stages of the light curve evolution of {\name} (see Figure \ref{fig:fig1}), translated to the rest frame, are shown as black vertical lines \revone{at the top}. For comparison, the evolution of the same quantity is shown for the three ASAS-SN TDEs \revone{(stars, see \citealt{holoien16b}), and for a sample of SLSNe-I with late-time observations (diamonds). For all the transients, the data are shown in the rest frame with respect to either the maximum of the light curve or the discovery date}.}
\label{fig:fig5}
\end{figure}

\begin{figure}
\centering
\subfloat{{\includegraphics[width=1.0\linewidth]{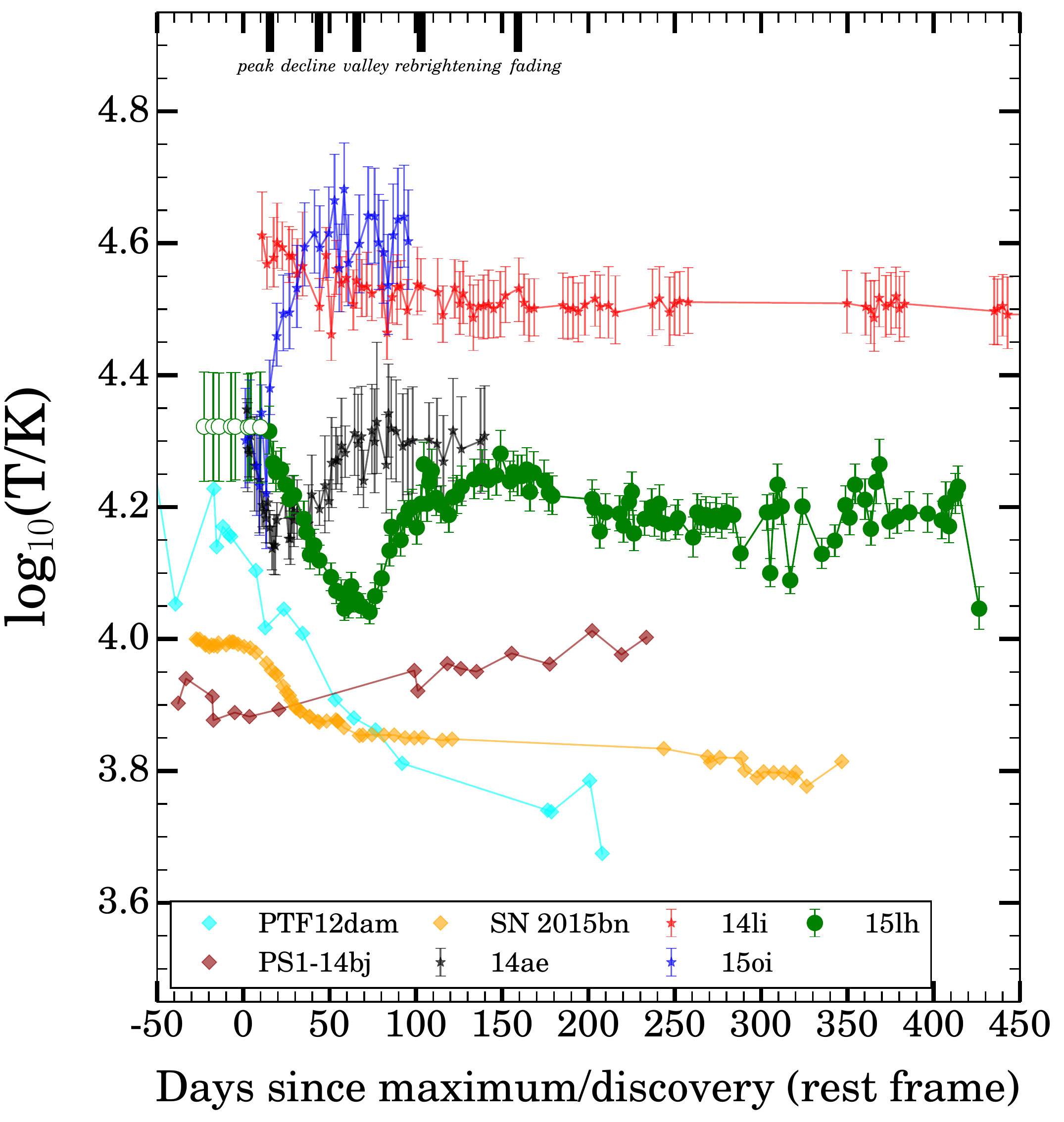}}}
\caption{Evolution of the effective photospheric temperatures of {\name}, the three ASAS-SN TDEs\revone{, and a sample of SLSNe-I with late-time observations}. The labeling and time-axis are the same as in Figure \ref{fig:fig5}.}
\label{fig:fig6}
\end{figure}

\begin{figure}
\centering
\subfloat{{\includegraphics[width=1.0\linewidth]{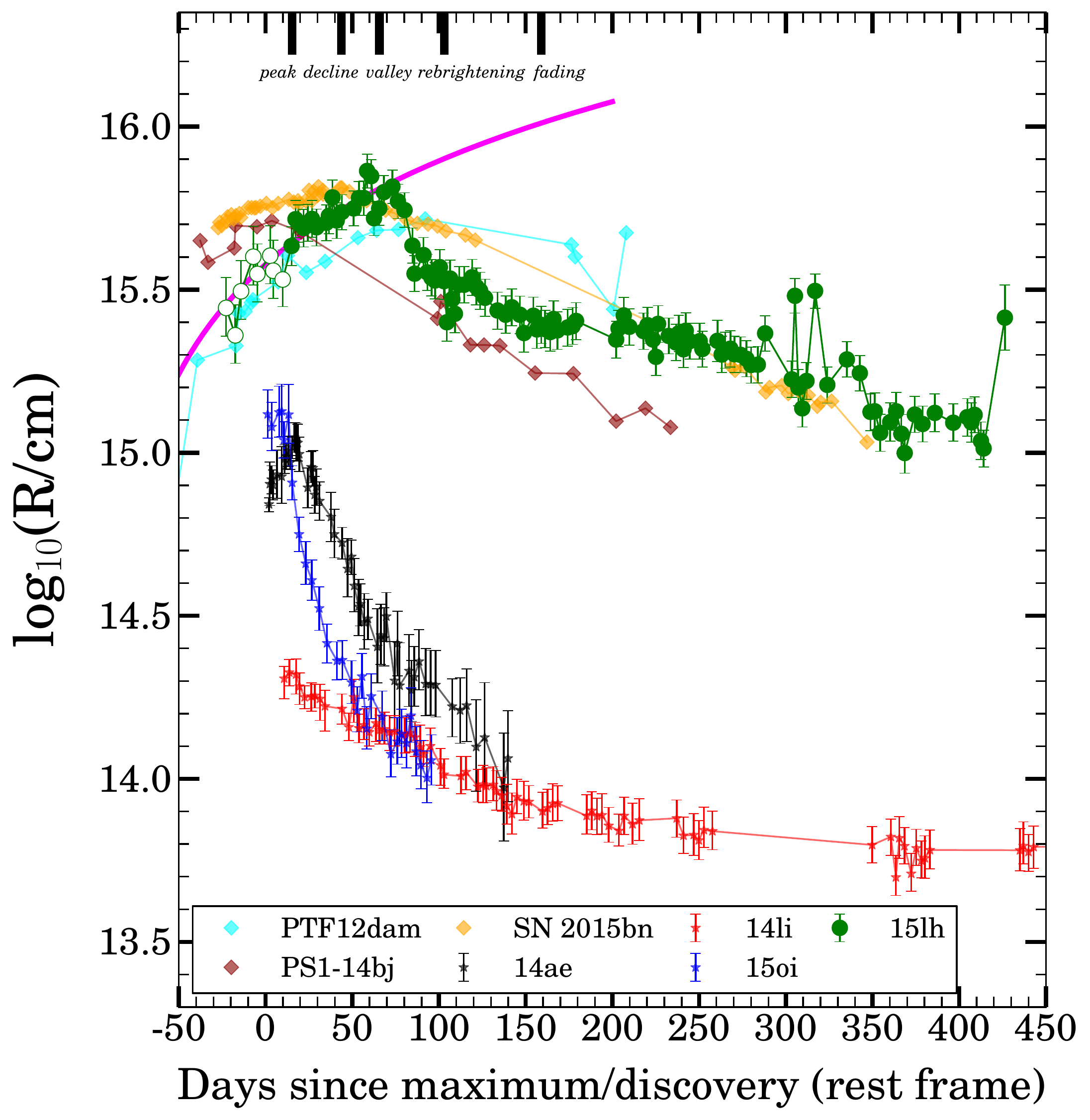}}}
\caption{Evolution of the apparent photospheric radii of {\name}, the three ASAS-SN TDEs\revone{, and a sample of SLSNe-I with late-time observations}. The \revone{purple} line shows the \revone{photospheric} radius evolution expected for a constant expansion at $\sim 4700$ km s$^{-1}$ for {\name}. The labeling and time-axis are otherwise the same as in Figures \ref{fig:fig5} and \ref{fig:fig6}.}
\label{fig:fig7}
\end{figure}

Figure \ref{fig:fig4} shows the observed and best-fit blackbody SED evolution, matching the five stages highlighted before. Models with the constant-temperature prior at early-times for the bolometric luminosity, effective temperature, and apparent photospheric radius are shown as the green symbols in Figures \ref{fig:fig5}, \ref{fig:fig6}, and \ref{fig:fig7}, respectively. For comparison and discussion (see \revone{\S}\ref{sec:discussion3}), the same quantities are also shown for the three ASAS-SN \revone{TDEs \citep{holoien14,holoien16b,holoien16a,brown16a,brown16b} and for a sample of SLSNe-I with available late-time data}.

The effects of the UV rebrightening are visible in Figure \ref{fig:fig4}. From the {\it peak} through the {\it decline} and into the {\it valley} phases, the SED of {\name} showed a monotonic decrease in luminosity while getting redder with a decreasing effective temperature. \revone{Later}, when the {\it rebrightening} occurred, \revone{these trends reversed}. The source increased \revone{both} in luminosity ($\sim 1.4$ times) and in effective temperature ($\sim 1.5$ times). Later on, for the {\it fading} stage, the SED was less luminous than \revone{during} the {\it rebrightening}, \revone{but it was still very blue and showed} an even higher effective temperature ($T_{\text{BB}}\simeq 18000$ K). \revone{This wavelength-dependent luminosity evolution shows that the ``additional'' radiation (in comparison with the monotonic SED decline at early-times) seen during the {\it rebrightening} is indeed peaking at the UV wavelengths.}

The evolution of the blackbody bolometric luminosity, effective temperature, and apparent photospheric radius of {\name} \revone{(}Figures \ref{fig:fig5}, \ref{fig:fig6}, and \ref{fig:fig7}\revone{)}, all show monotonic trends from the {\it peak} into the {\it valley}, with the photosphere cooling as it expands \revone{and becomes less luminous. During this time, the luminosity drops considerably, from $L_{\text{bol}} \simeq 6.2 \times 10^{11} L_{\odot}$ to $\simeq 1.1 \times 10^{11} L_{\odot}$, the effective temperature drops from $T_{\text{BB}} \simeq 20000$ K to $\simeq 11000$ K, and the photospheric radius increases from $R_{\text{BB}} \simeq 3 \times 10^{15}$ cm to $\simeq 7 \times 10^{15}$ cm (assuming the constant temperature prior at early-times).} The initial rapid rise in the radius of the apparent photosphere of {\name} roughly corresponds to a velocity of $\sim$ 4700 km s$^{-1}$, as indicated by the \revone{purple} line in Figure \ref{fig:fig7}.

Then, after $t \simeq 90$ days ($\simeq$ 73 days in the rest frame), when the UV rebrightening is seen in the multi-wavelength light curve (see Figure \ref{fig:fig1}), there is a rise in the effective temperature changing from $T_{\text{BB}} \simeq 11000$ K to $\simeq 18000$ K at $t \simeq 129$ days ($\simeq$ 105 days in the rest frame), and it maintains this high effective temperature even after entering the {\it fading} stage. After the {\it valley}, the luminosity shows a plateau at $L_{\text{bol}} \simeq 1.3 \times 10^{11} L_{\odot}$  for roughly 120 days ($\sim 97$ days in the rest frame). The \revone{largest increase}  in luminosity over this period is only of a factor of $\approx 1.4$. The large change in effective temperature at a nearly fixed luminosity then implies a significant drop in the apparent photospheric radius, from $R_{\text{BB}}\simeq 6.5 \times 10^{15}$ cm to $R_{\text{BB}}\simeq 2.3 \times 10^{15}$ cm. \revone{We} interpret this \revone{decrease in the photospheric radius} as seeing deeper into the ejecta, as the outer layers become optically thin.

\revone{After the {\it fading} stage, the bolometric luminosity decreases monotonically, the effective photospheric temperature oscillates with a constant overall trend, and the photospheric radius shows a decreasing trend. In the latest epochs, the luminosity has dropped to $L_{\text{bol}} \simeq 1.6 \times 10^{10} L_{\odot}$, while the photospheric radius has decreased to $R_{\text{BB}} \simeq 1.3 \times 10^{15}$ cm. At late-times the uncertainties in the blackbody parameters are greater because of the poorer photometry, as {\name} gets fainter.}

One possibility for explaining the behavior during the rebrightening phase is additional energy injection that both increased the effective temperature, and stabilized the bolometric luminosity. Alternatively, there could be two components to the luminosity: an optical component that continues to cool and fade with time, and a UV component that peaks at $t\simeq 150$ days ($\simeq$ 122 days in the rest frame). This second scenario would explain the change in effective temperature without needing additional late-time energy injection, but would likely require a change in opacity to allow more UV photons to escape at late-times (but see also \citealt{metzger14}).


\section{Discussion}
\label{sec:discussion}

\subsection{Energetics}
\label{sec:discussion1}

Blackbody models, while not perfect, are commonly used for SN SED fits (for recent examples see \citealt{arcavi16} and \citealt{nicholl16a}).\ Even though {\name} SEDs show deviations from our best-fit blackbody models (see Figure \ref{fig:fig4}), the approximation of its emission coming from a thermal photosphere allows us to model the bolometric luminosity and energy of the event, in addition to obtaining estimates for the effective temperature and photospheric radius. The X-ray non-detections derived in \revone{\S}\ref{sec:photometric_obs} \revone{(or the weak detection of \citealt{margutti16})} support a roughly thermal drop off in the SED at higher energies. 

Trapezoidal integration of the bolometric luminosity was used to calculate the energy emitted by {\name} from discovery through $t=\editfinal{\revone{525}}$ days after maximum. We estimate the total radiated energy \revone{(in the rest frame)} to be $E\simeq \editfinal{\revone{1.7}} \times 10^{52}$ ergs for the constant-temperature prior, and $E\simeq \editfinal{\revone{1.9}} \times 10^{52}$ ergs for the rising-temperature prior. The choice of the prior dominates the uncertainties in these estimates. 

To quantify the impact of the UV rebrightening in terms of energy, we separated the integral of the bolometric luminosity from discovery ($t \simeq -29$ days) to the {\it valley} stage ($t \simeq$ 81 days), and from the {\it valley} to the latest epoch ($t \simeq \editfinal{\revone{525}}$ days). {\name} radiated $\editfinal{\revone{\sim 9.6}} \times 10^{51}$ ergs (\editfinal{\revone{56}}\% of the total energy) before the rebrightening started, and $\editfinal{\revone{\sim 7.4}} \times 10^{51}$ ergs (\editfinal{\revone{44}}\%) after.

\subsection{{\name} and its UV rebrightening in the context of SLSNe powering mechanisms}
\label{sec:discussion2}

\revone{The physics behind SLSNe-I are poorly understood. Three commonly invoked mechanisms are: the radioactive decay of $^{56}$Ni \citep{gal-yam09,young10},  interaction between the SN ejecta and a dense circumstellar medium (CSM) \citep{chevalier11,ginzburg12,chatzopoulos13,sokorina16}, and the spindown of a magnetar (a highly magnetized and rapidly rotating neutron star) \citep{kasen10,woosley10,inserra13}.}

\revone{Given the extreme $\sim 2.2 \times 10^{45}$ ergs s$^{-1}$ peak bolometric luminosity, the radioactive decay of $^{56}$Ni is strongly disfavoured as one of the powering mechanisms for {\name}. Recently, \citet{kozyreva16} estimated that a nickel mass of 1500 $M_{\odot}$ would be required to explain this peak luminosity. However, the bolometric luminosity evolution obtained for this explosion differs significantly from the observed evolution (see their Figure 3). Consequently, this mechanism can be ruled out as a plausible driver of the explosion.}

\revone{For SLSNe-II (H-rich), which appear to be the high-luminosity counterparts of SNe IIn, CSM interaction seems to be the powering mechanism (\citealt{gal-yam12}, but see also \citealt{inserra16}). For this same mechanism to power SLSNe-I, however, the CSM would probably need to be H-poor, since no H emission lines are seen in the spectra of these events. This H-poor CSM interaction mechanism was favoured by \citet{chatzopoulos16} in order to explain the bolometric luminosity evolution of {\name}, although large ejecta and CSM masses ($\sim 36 M_{\odot}$ and $\sim 19.5 M_{\odot}$, respectively) would be needed. While providing a good fit to the overall luminosity evolution, detailed calculations of the expected multi-wavelength light curve are needed to see if the observed UV rebrightening can be explained in this context. More theoretical work is also needed to understand whether the absence of narrow emission lines in the spectra of {\name} disfavours this scenario.}

\revone{The magnetar model suggested by \citet{dong16} to explain the peak bolometric luminosity required a magnetar with a period $P \simeq 1$ ms and a magnetic field $B \simeq 10^{13} \textendash 10^{14}$ G, assuming efficient thermalization in the envelope, based on the models of \citet{kasen10}. These parameters could explain a radiated energy of up to $\sim 10^{52}$ ergs, but seemed extreme for a magnetar because they pushed the model close to the point where gravitational waves losses limit the available energy \citep{dong16}. Nevertheless, recent studies have claimed that the total radiated energy that can be extracted from the spinning down of such magnetar might be even higher. Both \citet{metzger15} and \citet{sukhbold16} argue that the upper energy limits of the magnetar model are still higher and are not exceeded by {\name}.} 

\revone{To reproduce the early-stages of the light curve of {\name} with the magnetar model, \citet{metzger15} used an ejecta mass of $M_{\text{ej}} = 3 M_{\odot}$. Given the ejecta mass, it is possible to make an order-of-magnitude estimate of the scale at which the ejecta should become optically thin, because the optical depth through the medium is
\begin{equation}
\tau \sim \kappa \Sigma \sim \kappa \frac{M_{\text{ej}}}{4 \pi R^2}. 
\end{equation}
Setting $\tau = 1$, and using an opacity of $0.2 \text{ cm}^2 \text{g}^{-1}$ \citep{metzger15} , we can solve for $R$ to find
\begin{equation}
R_{\tau=1} \sim  9.8 \times 10^{15} \text{ cm} \left( \frac{M_{\text{ej}}}{ 3 M_{\odot}}\right)^{1/2} \left( \frac{\kappa}{ 0.2 \text{ cm}^2 \text{g}^{-1}} \right)^{1/2}.
\end{equation}
This estimate is in rough agreement with our derived blackbody radius at the time the ejecta begins to recede, $R_{\text{bb}} \simeq 7 \times 10^{15}$ cm at $\sim 70$ days. Even for higher ejecta masses ($M_{\text{ej}} \simeq 11.2 M_{\odot}$, \citealt{sukhbold16}), or slightly different opacity values, the radius estimates are still of the same order.}

\revone{The standard magnetar model predicts a power-law luminosity decay $L_{\text{bol}} \propto t^{-2}$ \citep{inserra13}. This monotonic decay is obviously absent for {\name}, given that the observed UV rebrightening is associated with a plateau phase in the bolometric luminosity. A scenario where multiple processes are involved in powering SLSNe is certainly conceivable, and observational evidence for this has been found previously. An example is iPTF13ehe \citep{yan15,wang16}, a SLSN-I that showed a late ($\sim$ 250 days in the rest frame, \citealt{yan15}) $r$-band rebrightening, together with the appearance of broad H$\alpha$ emission in the spectrum. \citet{yan15} argued that the H$\alpha$ feature is caused by interaction between the SN ejecta and a H-rich CSM. Moreover, \citet{wang16} argued that all three of the possible SLSN mechanisms we discussed above could be involved in explaining the observed spectra and light curve evolution of iPTF13ehe. A magnetar and the decay of $^{56}$Ni could be simultaneously powering the early-time light curve, while interactions between the SN ejecta and the CSM would be the power source at late-times.}

\revone{Similarly, multiple power sources could also be powering {\name}, as suggested by \citet{chatzopoulos16}. {\name}, however, differs from iPTF13ehe. The rebrightenings were seen at different rest-frame wavelengths ($\lambda_{\text{rest}}\simeq 1600$\AA\ for {\name} and $\lambda_{\text{rest}} \simeq 4600$ \AA\ for iPTF13ehe), and {\name} shows no new prominent spectral features. It basically has a blue and mostly featureless spectrum at all times, with the hint of H$\alpha$ being associated with star formation in the host \citep{margutti16}. The $r$-band rebrightening seen in iPTF13ehe is quite different because it was accompanied by the appearance of strong H$\alpha$ emission.}   

\subsection{Comparison of {\name} with TDEs and SLSNe-I}
\label{sec:discussion3}

\subsubsection{Blackbody Parameters}
\label{sec:discussion3_bb}

\revone{In \citet{dong16}, the early-time observations of {\name} lead us to classify the transient as a SLSN-I, given the resemblance of its spectra with such transients.} Nevertheless, given that the position of {\name} is consistent with the center of its host galaxy (angular separation of $\sim$ 0\farcs2, implying a projected physical separation of $\sim$ 750 pc, \citealt{dong16}), a potential alternative interpretation for its nature is a TDE.

\revone{We compiled a list of both SLSNe-I and optically bright TDEs, preferably with late-time photometric observations, and Figures \ref{fig:fig5}, \ref{fig:fig6}, and \ref{fig:fig7} compare their rest-frame evolution in luminosity, temperature, and photospheric radius to {\name}. For the TDEs, we compare with the well studied systems ASASSN-14ae \citep{holoien14,brown16a}, ASASSN-14li \citep{holoien16a,brown16b}, and ASASSN-15oi \citep{holoien16b}. For the SLSNe-I, we compare with PTF12-dam \citep{nicholl13,chen15}, PS1-14bj \citep{lunnan16}, and SN 2015bn \citep{nicholl16a,nicholl16b}.}

\revone{PS1-14bj (at $z=0.5215$), PTF12dam (at $z=0.107$), and SN 2015bn  (at $z=0.1136$) are all SLSNe-I that evolved slowly, and could therefore be tracked until very late-times. PS1-14bj was followed-up in the optical but not in the UV, while PTF12dam and SN 2015bn were observed by {\swift} in the UV, but only until 22 days and 120 days after peak, respectively. For PS1-14bj, the blackbody parameters were taken from \citet{lunnan16} (derived using the $griz$-bands data). For PTF12dam we show the blackbody parameters from \citet{nicholl13}. For SN 2015bn, the blackbody parameters are from \citet{nicholl16a} (derived using the $UVW2$ to $K$-bands fits) until 140 days, and those of \citet{nicholl16b} (derived using $u$ to $K$-bands fits) for later times. Of these SLSNe-I, both PS1-14bj and SN 2015bn showed undulations in their multi-wavelength light curves.}

\revone{In Figure \ref{fig:fig5} we compare the bolometric luminosity evolution. {\name} stands out from both the TDEs and SLSNe-I, as it is $\sim 1$ order of magnitude more luminous than all comparison objects at almost all epochs. Excluding {\name}, both classes of transients can have similar peak bolometric luminosities, between $\sim 1 \textendash 7 \times 10^{10} L_{\odot}$. The ASAS-SN TDEs and most of the SLSNe-I have monotonically dropping luminosities. Only SN 2015bn shows a short plateau (15-day long at $L \simeq 2.5 \times 10^{10} L_{\odot}$), which was also much more pronounced in the UV than in the optical \citep{nicholl16a}, resembling {\name}.} 

\revone{Figure \ref{fig:fig6} shows a wide variety of behaviours for the effective temperature of the transients. There is a resemblance between the evolution of {\name} and ASASSN-14ae. Besides overlapping temperature values ($ 10000 \text{ K}\gtrsim T_{\text{BB}} \gtrsim 20000 \text{ K}$), they both decrease in temperature initially, followed by an increase. The timescales, however, are different. While the temperature of {\name} starts increasing at $t \simeq 70$ days, the temperature of ASASSN-14ae starts increasing at $t \simeq 20$ days.}

\revone{For SN 2015bn, there is a break in the temperature evolution coincident with the start of the luminosity plateau. It goes from a steady decrease in temperature ($\sim$ 11000 K at $t \simeq 0$ days), to a 15-day long plateau at $T_{\text{BB}} \simeq$ 8500 K. After this, the effective temperature remains almost constant at $\sim 7000$ K until very late-times (see also Figure 1 of \citealt{nicholl16b}). PS1-14bj has a very different evolution, showing an effective temperature that increases from $\sim$7500 K to $\sim$10000 K (even higher temperatures were obtained by \citet{lunnan16} using the $riz$-only fits). No clear distinction can be made between TDEs and SLSNe-I based on their effective temperature evolution.}

\revone{Figure \ref{fig:fig7} shows that there is a clear distinction between TDEs and SLSNe-I in their photospheric radius evolution. At time of maximum/discovery, the SLSNe-I tend to have larger photospheres than TDEs. This difference increases with time because the photospheric radii of the SLSNe-I shrink much more slowly than those of TDEs. Here, {\name} is clearly similar to the SLSNe-I and different from the TDEs, as it remains above $\sim 10^{15}$ cm for more than $\sim 450$ days. In fact, blackbody radii of $\sim 10^{15}$ cm are also typical of other SLSNe-I (e.g. \citealt{quimby11}). Figure \ref{fig:fig7} shows that ASASSN-14ae and SN 2015bn also have breaks in their photospheric radius evolution. However, once ASASSN-14ae starts shrinking, it evolves much faster. For SN 2015bn the radius evolution break happens simultaneously with the luminosity and temperature plateaus.}

\revone{Of the TDEs ASASSN-14ae would be the most similar to {\name} in temperature and radius evolution, although it is also the least luminous of the three. For the SLSNe-I, SN 2015bn shows a luminosity plateau, and the multi-wavelength light curve at this time has a similar behaviour to what was seen in the $U$-band of {\name} after the {\it valley}.}

\revone{It is important to note that SLSNe have not been generally observed at UV wavelengths at late-times. When they have been, as in SN 2015bn, the cadence tends to be poor. Figure \ref{fig:fig1} shows how important UV observations are for energetic transients, as the UV rebrightening of {\name} would have been missed without {\swift} observations. Only the $U$-band showed a strong change in its evolution, while the $V$ and $B$-bands simply continued to monotonically decline. Therefore, while the rebrightening seen in {\name} seems unique, we do not actually have the data to place strong constrains on the existence of similar episodes in SLSNe at late-times. This is different from the ASAS-SN TDEs, where the late-time UV data are available.}

\subsubsection{Spectroscopic Evolution}
\label{sec:discussion3_spec}

\revone{The spectroscopic features shown by {\name} are peculiar among SLSNe, but they are also peculiar among TDEs. The early-time spectroscopy, and in particular the resemblance of the absorption feature at $\sim$ 4100 \AA\ to SN 2010gx, pointed towards the correct redshift of {\name}, and was later confirmed by detecting the \ion{Mg}{2} absorption doublet at 2796 \AA\ and 2803 \AA\ \citep{dong16}.}

\revone{\citet{leloudas16} agree that the $\sim 4100$ \AA\ feature is not seen in TDEs, but argue that the absence of the $\sim 4400$ \AA\ feature in the early spectra of {\name} points towards a different physical origin than the \ion{O}{2} interpretation for SLSNe-I. In addition, the early-time spectrum of SN 2010gx was reproduced with simple assumptions by \citet{margutti16} (see their Figure 6), but this was not the case for {\name}, because of the missing $\sim 4400$ \AA\ absorption feature. The possibility that our initial conjecture of the redshift of {\name}, based on the similarities of its early spectra with other SLSNe-I, was only serendipitous, cannot be discarded.}

\revone{Looking at the overall evolution, Figure \ref{fig:fig2} shows that the spectrum {\name} is always continuum dominated, even at very late-times, without the appearance of nebular features. This differs from what has been seen in SLSNe-I and optically bright TDEs. These TDEs all show prominent He or H+He emission lines soon after peak (e.g., \citealt{arcavi14,holoien16a,holoien16b}), while the SLSNe-I stop being continuum dominated well before the time of our latest {\name} spectrum \citep{nicholl16a,lunnan16}. Although here we do not perform direct comparisons between the spectra of {\name} with TDEs and SLSNe, recently \citet{leloudas16} carried out such exercise (see their Supplementary Figure 1), showing that similarities and differences arise with both classes of transients, without finding any definitive evidence favouring either class based on their spectra.}

\subsubsection{Rates and Energies}
\label{sec:discussion4}

\begin{figure}
\centering
\subfloat{{\includegraphics[width=1.0\linewidth]{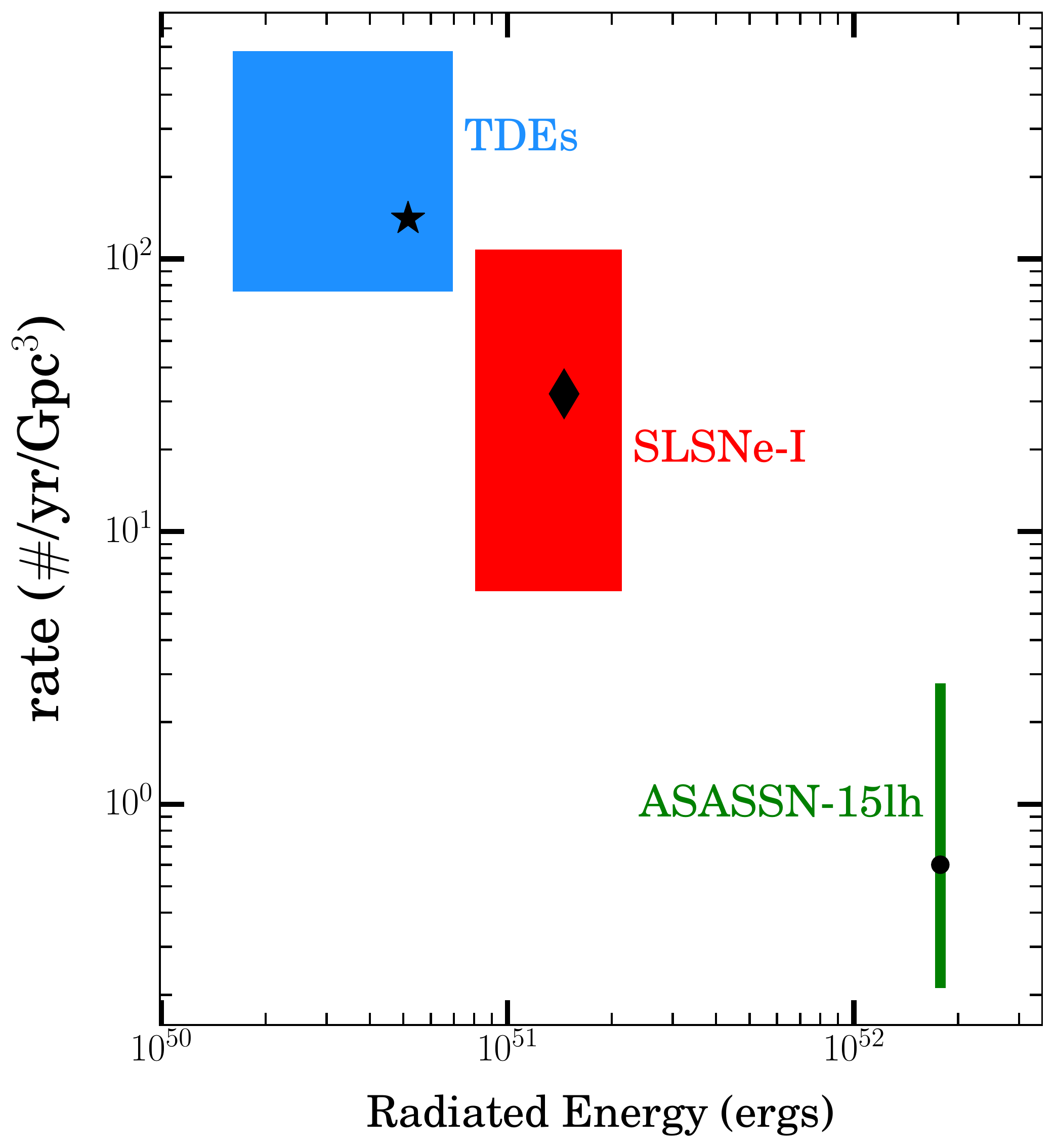}}}
\caption{\revone{Volumetric rate of events as a function of radiated energy for {\name} (green), SLSNe-I (red), and TDEs (blue). The typical values are indicated by the black symbols, while the surrounding regions show typical parameters range for each class.}}
\label{fig:fig8}
\end{figure}

\revone{As a last comparison, Figure \ref{fig:fig8} shows the rough distribution of TDEs and SLSNe-I as a function of rate and radiated energy. \citet{dong16} estimated the rate of {\name}-like events to be $r=0.60$ yr$^{-1}$ Gpc$^{-3}$ ($0.21 < r < 2.8$). For SLSNe-I, \citet{quimby13} estimated the rate to be $r=32$ yr$^{-1}$ Gpc$^{-3}$ ($6 < r < 109$) (see also \citealt{prajs16} for a rate at $z \sim 1$). For TDEs, \citet{vanvelzen14} estimated the rate to be $r = (1.5-2.0)_{-1.3}^{+2.7} \times 10^{-5}$ yr$^{-1}$ galaxy$^{-1}$, corresponding to a volumetric rate of $r=(4-8) \times 10^{1 \pm 0.4}$ yr$^{-1}$ Gpc$^{-3}$. The TDE rate estimated by \citet{holoien16a}, $r \simeq 4.1 \times 10^{-5}$ yr$^{-1}$ galaxy$^{-1}$ ($2.2 < r < 17.0$), is higher than that of \citet{vanvelzen14}, though they are consistent within the uncertainties. The discovery of the third ASAS-SN TDE promptly after the second one lead us to favour the rate estimated by \citet{holoien16a}. In order to translate this rate per galaxy to a volumetric one, we have followed \citet{kochanek16} and scaled it with a constant rate ratio relative to that of \citet{vanvelzen14}.}

\revone{We integrated the bolometric luminosity of the objects used in Figures \ref{fig:fig5}$\textendash$\ref{fig:fig7} using the trapezoidal rule to estimate the radiated energies. For SLSNe-I and TDEs, we have defined the minimum (maximum) of the energy interval to be the energy of the least (most) luminous event within the class, and the representative energy as the average energy of the members. For {\name}, we have defined the energy interval using the constant and rising temperature prior at early-times, and the representative energy as the average of those. For ASASSN-14li we adopted the total radiated energy of $E \simeq 7 \times 10^{50}$ ergs including the X-ray emission reported by \citet{brown16b}.}

\revone{Figure \ref{fig:fig8} shows that events as energetic as {\name} are rare among both TDEs and previously identified SLSNe-I, but the stretch from the SLSNe-I to {\name} is much less extreme than for the TDEs. It is plausible that {\name} is simply an extension of the SLSN distribution as we noted in \citet{dong16}. Making {\name} an extension of the TDE distribution is much more challenging because of the high TDE rates and the absence of TDEs of intermediate energies, in addition to the extreme parameters required under this scenario (see \citealt{leloudas16}).}

\section{Summary}
\label{sec:summary}

\revone{Here we report on the peculiar evolution of {\name}. The two most salient features are an extended UV rebrightening phase followed by a renewed fading, and a continued lack of spectroscopic features. The total radiated energy is roughly $E \simeq \editfinal{\revone{1.7}} \times 10^{52}$ ergs, with $\sim \editfinal{\revone{56}} \%$ coming from the initial phase and $\sim \editfinal{\revone{44}} \%$ from the later plateau and fading phases.}

\revone{We continue to favor the SLSN-I interpretation of this event over the primary alternative of a TDE. No optically bright TDE to date shows an absence of broad emission lines like {\name}. The recent results from \citet{margutti16} clearly show that the weak feature we discussed relative to our SOAR {\it fading} stage spectrum, and the feature interpreted as broad H$\alpha$ by \citet{leloudas16}, is actually a blend of narrow H$\alpha$ and [\ion{N}{2}] emission, probably due to star formation in the galaxy nucleus.}

\revone{We also compare the evolution of {\name} in luminosity, effective temperature, and photospheric radius to a selection of SLSNe-I and optically bright TDEs. Only the photospheric radius seems to clearly separate the two classes, and for this parameter {\name} behaves like a SLSN-I. The luminosity and temperature evolution provide no strong discrimination between the two classes, possibly due to the lack of good UV data for typical SLSNe. We also note that the radiated energy and the event rate implied by {\name} are easier to reconcile with the properties of SLSNe-I than with those of TDEs.}

\revone{Continued monitoring of {\name} will help to address some of these issues. For example, the X-ray emission detected by \citet{margutti16} is really only interpretable if it is seen to vary. The most likely route to resolving the riddle of {\name} is simply to obtain better population statistics. In particular, larger TDE samples will clarify the range of their evolutionary and spectral properties. Continuous UV observations of SLSNe will clarify whether features like the rebrightening and plateau phases of {\name} are ubiquitous in these events, as potentially hinted by SN 2015bn.}


\section*{Acknowledgments}

We thank Barry Madore and Juan Carlos Mateos Munoz for \revone{obtaining spectra}. \revone{We thank Matt Nicholl, Cosimo Inserra, Andrea Pastorello, and Ragnhild Lunnan for providing the blackbody parameters of SLSNe, and John Brown for providing the blackbody parameters of ASASSN-14ae. We thank Tuguldur Sukhbold and Tyler Holland-Ashford for useful discussions.}

The authors thank LCOGT and its staff for their continued support of ASAS-SN.

We thank Neil Gehrels for approving our various {\swift} ToO requests and the {\swift} science operation team for performing the observations. This research has made use of the XRT Data Analysis Software (XRTDAS) developed under the responsibility of the ASI Science Data Center (ASDC), Italy. This research has made use of data obtained through the High Energy  Astrophysics Science Archive Research Center Online Service, provided by the  NASA/Goddard Space Flight Center.

ASAS-SN is supported by NSF grant AST-1515927. Development of ASAS-SN has been supported by NSF grant AST-0908816, the Center for Cosmology and AstroParticle Physics at the Ohio State University, the Mt. Cuba Astronomical Foundation, and by George Skestos.

DGR is supported by The Ohio State University through the Dean's Distinguished University Fellowship. 
PC and SD are supported by ``the Strategic Priority Research Program-The Emergence of Cosmological Structures" of the Chinese Academy of Sciences (Grant No. XDB09000000) and Project 11573003 supported by NSFC.This research uses data obtained through the Telescope Access Program (TAP), which has been funded by ``the Strategic Priority Research Program- The Emergence of Cosmological Structures" of the Chinese Academy of Sciences (Grant No.11 XDB09000000) and the Special Fund for Astronomy from the Ministry of Finance. 
Support for JLP is in part provided by FONDECYT through the grant 1151445 and by the Ministry of Economy, Development, and Tourism's Millennium Science Initiative through grant IC120009, awarded to The Millennium Institute of Astrophysics, MAS. 
BJS is supported by NASA through Hubble Fellowship grant HF-51348.001 awarded by the Space Telescope Science Institute, which is operated by the Association of Universities for Research in Astronomy, Inc., for NASA, under contract NAS 5-26555. 
The UCSC group is supported in part by NSF grant AST-1518052 and from
fellowships from the Alfred P.\ Sloan Foundation and the David and
Lucile Packard Foundation to R.J.F.
TW-SH is supported by the DOE Computational Science Graduate Fellowship, grant number DE-FG02- 97ER25308. 
JFB is supported by NSF grant PHY-1404311.

This paper includes data gathered with the 6.5 meter Magellan Telescopes located at Las Campanas Observatory, Chile.


\bibliographystyle{mn2e}

\bibliography{bibliography}

\begin{thebibliography}{69}
\expandafter\ifx\csname natexlab\endcsname\relax\def\natexlab#1{#1}\fi

\bibitem[{{Arcavi} {et~al}\mbox{.}(2014){Arcavi}, {Gal-Yam}, {Sullivan}, {Pan},
  {Cenko}, {Horesh}, {Ofek}, {De Cia}, {Yan}, {Yang}, {Howell}, {Tal},
  {Kulkarni}, {Tendulkar}, {Tang}, {Xu}, {Sternberg}, {Cohen}, {Bloom},
  {Nugent}, {Kasliwal}, {Perley}, {Quimby}, {Miller}, {Theissen}, \&
  {Laher}}]{arcavi14}
{Arcavi} I. {et~al.}, 2014, \apj, 793, 38

\bibitem[{{Arcavi} {et~al}\mbox{.}(2016){Arcavi}, {Wolf}, {Howell}, {Bildsten},
  {Leloudas}, {Hardin}, {Prajs}, {Perley}, {Svirski}, {Gal-Yam}, {Katz},
  {McCully}, {Cenko}, {Lidman}, {Sullivan}, {Valenti}, {Astier}, {Balland},
  {Carlberg}, {Conley}, {Fouchez}, {Guy}, {Pain}, {Palanque-Delabrouille},
  {Perrett}, {Pritchet}, {Regnault}, {Rich}, \& {Ruhlmann-Kleider}}]{arcavi16}
{Arcavi} I. {et~al.}, 2016, \apj, 819, 35

\bibitem[{{Breeveld} {et~al}\mbox{.}(2010){Breeveld}, {Curran}, {Hoversten},
  {Koch}, {Landsman}, {Marshall}, {Page}, {Poole}, {Roming}, {Smith}, {Still},
  {Yershov}, {Blustin}, {Brown}, {Gronwall}, {Holland}, {Kuin}, {McGowan},
  {Rosen}, {Boyd}, {Broos}, {Carter}, {Chester}, {Hancock}, {Huckle}, {Immler},
  {Ivanushkina}, {Kennedy}, {Mason}, {Morgan}, {Oates}, {de Pasquale},
  {Schady}, {Siegel}, \& {vanden Berk}}]{breeveld10}
{Breeveld} A.~A. {et~al.}, 2010, \mnras, 406, 1687

\bibitem[{{Brown} {et~al}\mbox{.}(2016{\natexlab{a}}){Brown}, {Shappee},
  {Holoien}, {Stanek}, {Kochanek}, \& {Prieto}}]{brown16a}
{Brown} J.~S., {Shappee} B.~J., {Holoien} T.~W.-S., {Stanek} K.~Z., {Kochanek}
  C.~S., {Prieto} J.~L., 2016{\natexlab{a}}, \mnras, 462, 3993

\bibitem[{{Brown} {et~al}\mbox{.}(2016{\natexlab{b}}){Brown}, {W.-S Holoien},
  {Auchettl}, {Stanek}, {Kochanek}, {Shappee}, {Prieto}, \& {Grupe}}]{brown16b}
{Brown} J.~S., {W.-S Holoien} T., {Auchettl} K., {Stanek} K.~Z., {Kochanek}
  C.~S., {Shappee} B.~J., {Prieto} J.~L., {Grupe} D., 2016{\natexlab{b}}, ArXiv
  e-prints, arXiv:1609.04403

\bibitem[{{Brown}(2015)}]{brown15atel}
{Brown} P.~J., 2015, The Astronomer's Telegram, 8086

\bibitem[{{Brown} {et~al}\mbox{.}(2016{\natexlab{c}}){Brown}, {Yang}, {Cooke},
  {Olaes}, {Quimby}, {Baade}, {Gehrels}, {Hoeflich}, {Maund}, {Mould}, {Wang},
  \& {Wheeler}}]{brown16asassn15lh}
{Brown} P.~J. {et~al.}, 2016{\natexlab{c}}, \apj, 828, 3

\bibitem[{{Brown} {et~al}\mbox{.}(2013){Brown}, {Baliber}, {Bianco}, {Bowman},
  {Burleson}, {Conway}, {Crellin}, {Depagne}, {De Vera}, {Dilday}, {Dragomir},
  {Dubberley}, {Eastman}, {Elphick}, {Falarski}, {Foale}, {Ford}, {Fulton},
  {Garza}, {Gomez}, {Graham}, {Greene}, {Haldeman}, {Hawkins}, {Haworth},
  {Haynes}, {Hidas}, {Hjelstrom}, {Howell}, {Hygelund}, {Lister}, {Lobdill},
  {Martinez}, {Mullins}, {Norbury}, {Parrent}, {Paulson}, {Petry}, {Pickles},
  {Posner}, {Rosing}, {Ross}, {Sand}, {Saunders}, {Shobbrook}, {Shporer},
  {Street}, {Thomas}, {Tsapras}, {Tufts}, {Valenti}, {Vander Horst}, {Walker},
  {White}, \& {Willis}}]{brown13}
{Brown} T.~M. {et~al.}, 2013, \pasp, 125, 1031

\bibitem[{{Burrows} {et~al}\mbox{.}(2005){Burrows}, {Hill}, {Nousek}, {Kennea},
  {Wells}, {Osborne}, {Abbey}, {Beardmore}, {Mukerjee}, {Short}, {Chincarini},
  {Campana}, {Citterio}, {Moretti}, {Pagani}, {Tagliaferri}, {Giommi},
  {Capalbi}, {Tamburelli}, {Angelini}, {Cusumano}, {Br{\"a}uninger}, {Burkert},
  \& {Hartner}}]{burrows05}
{Burrows} D.~N. {et~al.}, 2005, Space Sci. Rev., 120, 165

\bibitem[{{Chatzopoulos} {et~al}\mbox{.}(2013){Chatzopoulos}, {Wheeler},
  {Vinko}, {Horvath}, \& {Nagy}}]{chatzopoulos13}
{Chatzopoulos} E., {Wheeler} J.~C., {Vinko} J., {Horvath} Z.~L., {Nagy} A.,
  2013, \apj, 773, 76

\bibitem[{{Chatzopoulos} {et~al}\mbox{.}(2016){Chatzopoulos}, {Wheeler},
  {Vinko}, {Nagy}, {Wiggins}, \& {Even}}]{chatzopoulos16}
{Chatzopoulos} E., {Wheeler} J.~C., {Vinko} J., {Nagy} A.~P., {Wiggins} B.~K.,
  {Even} W.~P., 2016, \apj, 828, 94

\bibitem[{{Chen} {et~al}\mbox{.}(2015){Chen}, {Smartt}, {Jerkstrand},
  {Nicholl}, {Bresolin}, {Kotak}, {Polshaw}, {Rest}, {Kudritzki}, {Zheng},
  {Elias-Rosa}, {Smith}, {Inserra}, {Wright}, {Kankare}, {Kangas}, \&
  {Fraser}}]{chen15}
{Chen} T.-W. {et~al.}, 2015, \mnras, 452, 1567

\bibitem[{{Chevalier} \& {Irwin}(2011)}]{chevalier11}
{Chevalier} R.~A., {Irwin} C.~M., 2011, \apjl, 729, L6

\bibitem[{{Clemens}, {Crain} \& {Anderson}(2004){Clemens}, {Crain}, \&
  {Anderson}}]{clemens04}
{Clemens} J.~C., {Crain} J.~A., {Anderson} R., 2004, in Society of
  Photo-Optical Instrumentation Engineers (SPIE) Conference Series, Vol. 5492,
  Ground-based Instrumentation for Astronomy, {Moorwood} A.~F.~M., {Iye} M.,
  eds., pp. 331--340

\bibitem[{{Crawford} {et~al}\mbox{.}(2010){Crawford}, {Still}, {Schellart},
  {Balona}, {Buckley}, {Dugmore}, {Gulbis}, {Kniazev}, {Kotze}, {Loaring},
  {Nordsieck}, {Pickering}, {Potter}, {Romero Colmenero}, {Vaisanen},
  {Williams}, \& {Zietsman}}]{crawford10}
{Crawford} S.~M. {et~al.}, 2010, in Society of Photo-Optical Instrumentation
  Engineers (SPIE) Conference Series, Vol. 7737, Society of Photo-Optical
  Instrumentation Engineers (SPIE) Conference Series, p.~25

\bibitem[{{Dong} {et~al}\mbox{.}(2016){Dong}, {Shappee}, {Prieto}, {Jha},
  {Stanek}, {Holoien}, {Kochanek}, {Thompson}, {Morrell}, {Thompson}, {Basu},
  {Beacom}, {Bersier}, {Brimacombe}, {Brown}, {Bufano}, {Chen}, {Conseil},
  {Danilet}, {Falco}, {Grupe}, {Kiyota}, {Masi}, {Nicholls}, {Olivares E.},
  {Pignata}, {Pojmanski}, {Simonian}, {Szczygiel}, \& {Wo{\'z}niak}}]{dong16}
{Dong} S. {et~al.}, 2016, Science, 351, 257

\bibitem[{{Foley} {et~al}\mbox{.}(2003){Foley}, {Papenkova}, {Swift},
  {Filippenko}, {Li}, {Mazzali}, {Chornock}, {Leonard}, \& {Van Dyk}}]{foley03}
{Foley} R.~J. {et~al.}, 2003, \pasp, 115, 1220

\bibitem[{{Gal-Yam}(2012)}]{gal-yam12}
{Gal-Yam} A., 2012, Science, 337, 927

\bibitem[{{Gal-Yam} {et~al}\mbox{.}(2009){Gal-Yam}, {Mazzali}, {Ofek},
  {Nugent}, {Kulkarni}, {Kasliwal}, {Quimby}, {Filippenko}, {Cenko},
  {Chornock}, {Waldman}, {Kasen}, {Sullivan}, {Beshore}, {Drake}, {Thomas},
  {Bloom}, {Poznanski}, {Miller}, {Foley}, {Silverman}, {Arcavi}, {Ellis}, \&
  {Deng}}]{gal-yam09}
{Gal-Yam} A. {et~al.}, 2009, \nat, 462, 624

\bibitem[{{Ginzburg} \& {Balberg}(2012)}]{ginzburg12}
{Ginzburg} S., {Balberg} S., 2012, \apj, 757, 178

\bibitem[{{Henden} {et~al}\mbox{.}(2015){Henden}, {Levine}, {Terrell}, \&
  {Welch}}]{henden15}
{Henden} A.~A., {Levine} S., {Terrell} D., {Welch} D.~L., 2015, in American
  Astronomical Society Meeting Abstracts, Vol. 225, American Astronomical
  Society Meeting Abstracts, p. 336.16

\bibitem[{{Hill} {et~al}\mbox{.}(2004){Hill}, {Burrows}, {Nousek}, {Abbey},
  {Ambrosi}, {Br{\"a}uninger}, {Burkert}, {Campana}, {Cheruvu}, {Cusumano},
  {Freyberg}, {Hartner}, {Klar}, {Mangels}, {Moretti}, {Mori}, {Morris},
  {Short}, {Tagliaferri}, {Watson}, {Wood}, \& {Wells}}]{hill04}
{Hill} J.~E. {et~al.}, 2004, in SPIE Conference Series, Vol. 5165, X-Ray and
  Gamma-Ray Instrumentation for Astronomy XIII, {Flanagan} K.~A., {Siegmund}
  O.~H.~W., eds., pp. 217--231

\bibitem[{{Holoien} {et~al}\mbox{.}(2016{\natexlab{a}}){Holoien}, {Kochanek},
  {Prieto}, {Grupe}, {Chen}, {Godoy-Rivera}, {Stanek}, {Shappee}, {Dong},
  {Brown}, {Basu}, {Beacom}, {Bersier}, {Brimacombe}, {Carlson}, {Falco},
  {Johnston}, {Madore}, {Pojmanski}, \& {Seibert}}]{holoien16b}
{Holoien} T.~W.-S. {et~al.}, 2016{\natexlab{a}}, \mnras

\bibitem[{{Holoien} {et~al}\mbox{.}(2016{\natexlab{b}}){Holoien}, {Kochanek},
  {Prieto}, {Stanek}, {Dong}, {Shappee}, {Grupe}, {Brown}, {Basu}, {Beacom},
  {Bersier}, {Brimacombe}, {Danilet}, {Falco}, {Guo}, {Jose}, {Herczeg},
  {Long}, {Pojmanski}, {Simonian}, {Szczygie{\l}}, {Thompson}, {Thorstensen},
  {Wagner}, \& {Wo{\'z}niak}}]{holoien16a}
{Holoien} T.~W.-S. {et~al.}, 2016{\natexlab{b}}, \mnras, 455, 2918

\bibitem[{{Holoien} {et~al}\mbox{.}(2014){Holoien}, {Prieto}, {Bersier},
  {Kochanek}, {Stanek}, {Shappee}, {Grupe}, {Basu}, {Beacom}, {Brimacombe},
  {Brown}, {Davis}, {Jencson}, {Pojmanski}, \& {Szczygie{\l}}}]{holoien14}
{Holoien} T.~W.-S. {et~al.}, 2014, \mnras, 445, 3263

\bibitem[{{Horne}(1986)}]{horne86}
{Horne} K., 1986, \pasp, 98, 609

\bibitem[{{Inserra} {et~al}\mbox{.}(2016){Inserra}, {Smartt}, {Gall},
  {Leloudas}, {Chen}, {Schulze}, {Jerkstarnd}, {Nicholl}, {Anderson}, {Arcavi},
  {Benetti}, {Cartier}, {Childress}, {Della Valle}, {Flewelling}, {Fraser},
  {Gal-Yam}, {Gutierrez}, {Hosseinzadeh}, {Howell}, {Huber}, {Kankare},
  {Magnier}, {Maguire}, {McCully}, {Prajs}, {Primak}, {Scalzo}, {Schmidt},
  {Smith}, {Tucker}, {Valenti}, {Wilman}, {Young}, \& {Yuan}}]{inserra16}
{Inserra} C. {et~al.}, 2016, ArXiv e-prints, arXiv:1604.01226

\bibitem[{{Inserra} {et~al}\mbox{.}(2013){Inserra}, {Smartt}, {Jerkstrand},
  {Valenti}, {Fraser}, {Wright}, {Smith}, {Chen}, {Kotak}, {Pastorello},
  {Nicholl}, {Bresolin}, {Kudritzki}, {Benetti}, {Botticella}, {Burgett},
  {Chambers}, {Ergon}, {Flewelling}, {Fynbo}, {Geier}, {Hodapp}, {Howell},
  {Huber}, {Kaiser}, {Leloudas}, {Magill}, {Magnier}, {McCrum}, {Metcalfe},
  {Price}, {Rest}, {Sollerman}, {Sweeney}, {Taddia}, {Taubenberger}, {Tonry},
  {Wainscoat}, {Waters}, \& {Young}}]{inserra13}
{Inserra} C. {et~al.}, 2013, \apj, 770, 128

\bibitem[{{Kaiser} {et~al}\mbox{.}(2002){Kaiser}, {Aussel}, {Burke},
  {Boesgaard}, {Chambers}, {Chun}, {Heasley}, {Hodapp}, {Hunt}, {Jedicke},
  {Jewitt}, {Kudritzki}, {Luppino}, {Maberry}, {Magnier}, {Monet}, {Onaka},
  {Pickles}, {Rhoads}, {Simon}, {Szalay}, {Szapudi}, {Tholen}, {Tonry},
  {Waterson}, \& {Wick}}]{kaiser02}
{Kaiser} N. {et~al.}, 2002, in Proc. SPIE, Vol. 4836, Survey and Other
  Telescope Technologies and Discoveries, {Tyson} J.~A., {Wolff} S., eds., pp.
  154--164

\bibitem[{{Kasen} \& {Bildsten}(2010)}]{kasen10}
{Kasen} D., {Bildsten} L., 2010, \apj, 717, 245

\bibitem[{{Kochanek}(2016)}]{kochanek16}
{Kochanek} C.~S., 2016, \mnras, 461, 371

\bibitem[{{Kool} {et~al}\mbox{.}(2015){Kool}, {Ryder}, {Stockdale},
  {Romero-Canizales}, {Prieto}, \& {Kotak}}]{kool15atel}
{Kool} E.~C., {Ryder} S.~D., {Stockdale} C.~J., {Romero-Canizales} C., {Prieto}
  J.~L., {Kotak} R., 2015, The Astronomer's Telegram, 8388

\bibitem[{{Kozyreva} {et~al}\mbox{.}(2016){Kozyreva}, {Hirschi}, {Blinnikov},
  \& {den Hartogh}}]{kozyreva16}
{Kozyreva} A., {Hirschi} R., {Blinnikov} S., {den Hartogh} J., 2016, \mnras,
  459, L21

\bibitem[{{Kraft}, {Burrows} \& {Nousek}(1991){Kraft}, {Burrows}, \&
  {Nousek}}]{kraft91}
{Kraft} R.~P., {Burrows} D.~N., {Nousek} J.~A., 1991, \apj, 374, 344

\bibitem[{{Law} {et~al}\mbox{.}(2009){Law}, {Kulkarni}, {Dekany}, {Ofek},
  {Quimby}, {Nugent}, {Surace}, {Grillmair}, {Bloom}, {Kasliwal}, {Bildsten},
  {Brown}, {Cenko}, {Ciardi}, {Croner}, {Djorgovski}, {van Eyken},
  {Filippenko}, {Fox}, {Gal-Yam}, {Hale}, {Hamam}, {Helou}, {Henning},
  {Howell}, {Jacobsen}, {Laher}, {Mattingly}, {McKenna}, {Pickles},
  {Poznanski}, {Rahmer}, {Rau}, {Rosing}, {Shara}, {Smith}, {Starr},
  {Sullivan}, {Velur}, {Walters}, \& {Zolkower}}]{law09}
{Law} N.~M. {et~al.}, 2009, \pasp, 121, 1395

\bibitem[{{Leloudas} {et~al}\mbox{.}(2016){Leloudas}, {Fraser}, {Stone}, {van
  Velzen}, {Jonker}, {Arcavi}, {Fremling}, {Maund}, {Smartt}, {Kruhler},
  {Miller-Jones}, {Vreeswijk}, {Gal-Yam}, {Mazzali}, {De Cia}, {Howell},
  {Inserra}, {Patat}, {de Ugarte Postigo}, {Yaron}, {Ashall}, {Bar},
  {Campbell}, {Chen}, {Childress}, {Elias-Rosa}, {Harmanen}, {Hosseinzadeh},
  {Johansson}, {Kangas}, {Kankare}, {Kim}, {Kuncarayakti}, {Lyman}, {Magee},
  {Maguire}, {Malesani}, {Mattila}, {McCully}, {Nicholl}, {Prentice},
  {Romero-Canizales}, {Schulze}, {Smith}, {Sollerman}, {Sullivan}, {Tucker},
  {Valenti}, {Wheeler}, \& {Young}}]{leloudas16}
{Leloudas} G. {et~al.}, 2016, ArXiv e-prints, arXiv:1609.02927

\bibitem[{{Lunnan} {et~al}\mbox{.}(2016){Lunnan}, {Chornock}, {Berger},
  {Milisavljevic}, {Jones}, {Rest}, {Fong}, {Fransson}, {Margutti}, {Drout},
  {Blanchard}, {Challis}, {Cowperthwaite}, {Foley}, {Kirshner}, {Morrell},
  {Riess}, {Roth}, {Scolnic}, {Smartt}, {Smith}, {Villar}, {Chambers},
  {Draper}, {Huber}, {Kaiser}, {Kudritzki}, {Magnier}, {Metcalfe}, \&
  {Waters}}]{lunnan16}
{Lunnan} R. {et~al.}, 2016, \apj, 831, 144

\bibitem[{{Lunnan} {et~al}\mbox{.}(2015){Lunnan}, {Chornock}, {Berger}, {Rest},
  {Fong}, {Scolnic}, {Jones}, {Soderberg}, {Challis}, {Drout}, {Foley},
  {Huber}, {Kirshner}, {Leibler}, {Marion}, {McCrum}, {Milisavljevic},
  {Narayan}, {Sanders}, {Smartt}, {Smith}, {Tonry}, {Burgett}, {Chambers},
  {Flewelling}, {Kudritzki}, {Wainscoat}, \& {Waters}}]{lunnan15}
{Lunnan} R. {et~al.}, 2015, \apj, 804, 90

\bibitem[{{Maddox} {et~al}\mbox{.}(1990){Maddox}, {Efstathiou}, {Sutherland},
  \& {Loveday}}]{maddox90}
{Maddox} S.~J., {Efstathiou} G., {Sutherland} W.~J., {Loveday} J., 1990,
  \mnras, 243, 692

\bibitem[{{Margutti} {et~al}\mbox{.}(2016){Margutti}, {Metzger}, {Chornock},
  {Milisavljevic}, {Berger}, {Blanchard}, {Guidorzi}, {Migliori}, {Kamble},
  {Lunnan}, {Nicholl}, {Coppejans}, {Dall'Osso}, {Drout}, {Perna}, \&
  {Sbarufatti}}]{margutti16}
{Margutti} R. {et~al.}, 2016, ArXiv e-prints, arXiv:1610.01632

\bibitem[{{Marshall} {et~al}\mbox{.}(2008){Marshall}, {Burles}, {Thompson},
  {Shectman}, {Bigelow}, {Burley}, {Birk}, {Estrada}, {Jones}, {Smith},
  {Kowal}, {Castillo}, {Storts}, \& {Ortiz}}]{marshall08}
{Marshall} J.~L. {et~al.}, 2008, in Proc. SPIE, Vol. 7014, Ground-based and
  Airborne Instrumentation for Astronomy II, p. 701454

\bibitem[{{Metzger} {et~al}\mbox{.}(2015){Metzger}, {Margalit}, {Kasen}, \&
  {Quataert}}]{metzger15}
{Metzger} B.~D., {Margalit} B., {Kasen} D., {Quataert} E., 2015, \mnras, 454,
  3311

\bibitem[{{Metzger} {et~al}\mbox{.}(2014){Metzger}, {Vurm}, {Hasco{\"e}t}, \&
  {Beloborodov}}]{metzger14}
{Metzger} B.~D., {Vurm} I., {Hasco{\"e}t} R., {Beloborodov} A.~M., 2014,
  \mnras, 437, 703

\bibitem[{{Milisavljevic} {et~al}\mbox{.}(2015){Milisavljevic}, {James},
  {Marshall}, {Patnaude}, {Margutti}, {Parrent}, \&
  {Kamble}}]{milisavljevic15atel}
{Milisavljevic} D., {James} D.~J., {Marshall} J.~L., {Patnaude} D., {Margutti}
  R., {Parrent} J., {Kamble} A., 2015, The Astronomer's Telegram, 8216

\bibitem[{{Neill} {et~al}\mbox{.}(2011){Neill}, {Sullivan}, {Gal-Yam},
  {Quimby}, {Ofek}, {Wyder}, {Howell}, {Nugent}, {Seibert}, {Martin},
  {Overzier}, {Barlow}, {Foster}, {Friedman}, {Morrissey}, {Neff},
  {Schiminovich}, {Bianchi}, {Donas}, {Heckman}, {Lee}, {Madore}, {Milliard},
  {Rich}, \& {Szalay}}]{neill11}
{Neill} J.~D. {et~al.}, 2011, \apj, 727, 15

\bibitem[{{Nicholl} {et~al}\mbox{.}(2016{\natexlab{a}}){Nicholl}, {Berger},
  {Margutti}, {Chornock}, {Blanchard}, {Jerkstrand}, {Smartt}, {Arcavi},
  {Challis}, {Chambers}, {Chen}, {Cowperthwaite}, {Gal-Yam}, {Hosseinzadeh},
  {Howell}, {Inserra}, {Kankare}, {Magnier}, {Maguire}, {Mazzali}, {McCully},
  {Milisavljevic}, {Smith}, {Taubenberger}, {Valenti}, {Wainscoat}, {Yaron}, \&
  {Young}}]{nicholl16b}
{Nicholl} M. {et~al.}, 2016{\natexlab{a}}, \apjl, 828, L18

\bibitem[{{Nicholl} {et~al}\mbox{.}(2016{\natexlab{b}}){Nicholl}, {Berger},
  {Smartt}, {Margutti}, {Kamble}, {Alexander}, {Chen}, {Inserra}, {Arcavi},
  {Blanchard}, {Cartier}, {Chambers}, {Childress}, {Chornock}, {Cowperthwaite},
  {Drout}, {Flewelling}, {Fraser}, {Gal-Yam}, {Galbany}, {Harmanen}, {Holoien},
  {Hosseinzadeh}, {Howell}, {Huber}, {Jerkstrand}, {Kankare}, {Kochanek},
  {Lin}, {Lunnan}, {Magnier}, {Maguire}, {McCully}, {McDonald}, {Metzger},
  {Milisavljevic}, {Mitra}, {Reynolds}, {Saario}, {Shappee}, {Smith},
  {Valenti}, {Villar}, {Waters}, \& {Young}}]{nicholl16a}
{Nicholl} M. {et~al.}, 2016{\natexlab{b}}, \apj, 826, 39

\bibitem[{{Nicholl} {et~al}\mbox{.}(2013){Nicholl}, {Smartt}, {Jerkstrand},
  {Inserra}, {McCrum}, {Kotak}, {Fraser}, {Wright}, {Chen}, {Smith}, {Young},
  {Sim}, {Valenti}, {Howell}, {Bresolin}, {Kudritzki}, {Tonry}, {Huber},
  {Rest}, {Pastorello}, {Tomasella}, {Cappellaro}, {Benetti}, {Mattila},
  {Kankare}, {Kangas}, {Leloudas}, {Sollerman}, {Taddia}, {Berger}, {Chornock},
  {Narayan}, {Stubbs}, {Foley}, {Lunnan}, {Soderberg}, {Sanders},
  {Milisavljevic}, {Margutti}, {Kirshner}, {Elias-Rosa}, {Morales-Garoffolo},
  {Taubenberger}, {Botticella}, {Gezari}, {Urata}, {Rodney}, {Riess},
  {Scolnic}, {Wood-Vasey}, {Burgett}, {Chambers}, {Flewelling}, {Magnier},
  {Kaiser}, {Metcalfe}, {Morgan}, {Price}, {Sweeney}, \& {Waters}}]{nicholl13}
{Nicholl} M. {et~al.}, 2013, \nat, 502, 346

\bibitem[{{Pastorello} {et~al}\mbox{.}(2010){Pastorello}, {Smartt},
  {Botticella}, {Maguire}, {Fraser}, {Smith}, {Kotak}, {Magill}, {Valenti},
  {Young}, {Gezari}, {Bresolin}, {Kudritzki}, {Howell}, {Rest}, {Metcalfe},
  {Mattila}, {Kankare}, {Huang}, {Urata}, {Burgett}, {Chambers}, {Dombeck},
  {Flewelling}, {Grav}, {Heasley}, {Hodapp}, {Kaiser}, {Luppino}, {Lupton},
  {Magnier}, {Monet}, {Morgan}, {Onaka}, {Price}, {Rhoads}, {Siegmund},
  {Stubbs}, {Sweeney}, {Tonry}, {Wainscoat}, {Waterson}, {Waters}, \&
  {Wynn-Williams}}]{pastorello10}
{Pastorello} A. {et~al.}, 2010, \apjl, 724, L16

\bibitem[{{Poole} {et~al}\mbox{.}(2008){Poole}, {Breeveld}, {Page}, {Landsman},
  {Holland}, {Roming}, {Kuin}, {Brown}, {Gronwall}, {Hunsberger}, {Koch},
  {Mason}, {Schady}, {vanden Berk}, {Blustin}, {Boyd}, {Broos}, {Carter},
  {Chester}, {Cucchiara}, {Hancock}, {Huckle}, {Immler}, {Ivanushkina},
  {Kennedy}, {Marshall}, {Morgan}, {Pandey}, {de Pasquale}, {Smith}, \&
  {Still}}]{poole08}
{Poole} T.~S. {et~al.}, 2008, \mnras, 383, 627

\bibitem[{{Prajs} {et~al}\mbox{.}(2016){Prajs}, {Sullivan}, {Smith}, {Levan},
  {Karpenka}, {Edwards}, {Walker}, {Wolf}, {Balland}, {Carlberg}, {Howell},
  {Lidman}, {Pain}, {Pritchet}, \& {Ruhlmann-Kleider}}]{prajs16}
{Prajs} S. {et~al.}, 2016, \mnras

\bibitem[{{Quimby}(2006)}]{quimby06}
{Quimby} R.~M., 2006, PhD thesis, The University of Texas at Austin

\bibitem[{{Quimby} {et~al}\mbox{.}(2011){Quimby}, {Kulkarni}, {Kasliwal},
  {Gal-Yam}, {Arcavi}, {Sullivan}, {Nugent}, {Thomas}, {Howell}, E.,
  {Bildsten}, {Theissen}, {Law}, {Dekany}, {Rahmer}, {Hale}, {Smith}, {Ofek},
  {Zolkower}, {Velur}, {Walters}, {Henning}, {Bui}, {McKenna}, {Poznanski},
  {Cenko}, \& {Levitan}}]{quimby11}
{Quimby} R.~M. {et~al.}, 2011, \nat, 474, 487

\bibitem[{{Quimby} {et~al}\mbox{.}(2013){Quimby}, {Yuan}, {Akerlof}, \&
  {Wheeler}}]{quimby13}
{Quimby} R.~M., {Yuan} F., {Akerlof} C., {Wheeler} J.~C., 2013, \mnras, 431,
  912

\bibitem[{{Rau} {et~al}\mbox{.}(2009){Rau}, {Kulkarni}, {Law}, {Bloom},
  {Ciardi}, {Djorgovski}, {Fox}, {Gal-Yam}, {Grillmair}, {Kasliwal}, {Nugent},
  {Ofek}, {Quimby}, {Reach}, {Shara}, {Bildsten}, {Cenko}, {Drake},
  {Filippenko}, {Helfand}, {Helou}, {Howell}, {Poznanski}, \&
  {Sullivan}}]{rau09}
{Rau} A. {et~al.}, 2009, \pasp, 121, 1334

\bibitem[{{Roming} {et~al}\mbox{.}(2005){Roming}, {Kennedy}, {Mason}, {Nousek},
  {Ahr}, {Bingham}, {Broos}, {Carter}, {Hancock}, {Huckle}, {Hunsberger},
  {Kawakami}, {Killough}, {Koch}, {McLelland}, {Smith}, {Smith}, {Soto},
  {Boyd}, {Breeveld}, {Holland}, {Ivanushkina}, {Pryzby}, {Still}, \&
  {Stock}}]{roming05}
{Roming} P.~W.~A. {et~al.}, 2005, Space Sci. Rev., 120, 95

\bibitem[{{Schlafly} \& {Finkbeiner}(2011)}]{schlafly11}
{Schlafly} E.~F., {Finkbeiner} D.~P., 2011, \apj, 737, 103

\bibitem[{{Shappee} {et~al}\mbox{.}(2014){Shappee}, {Prieto}, {Grupe},
  {Kochanek}, {Stanek}, {De Rosa}, {Mathur}, {Zu}, {Peterson}, {Pogge},
  {Komossa}, {Im}, {Jencson}, {Holoien}, {Basu}, {Beacom}, {Szczygie{\l}},
  {Brimacombe}, {Adams}, {Campillay}, {Choi}, {Contreras}, {Dietrich},
  {Dubberley}, {Elphick}, {Foale}, {Giustini}, {Gonzalez}, {Hawkins}, {Howell},
  {Hsiao}, {Koss}, {Leighly}, {Morrell}, {Mudd}, {Mullins}, {Nugent},
  {Parrent}, {Phillips}, {Pojmanski}, {Rosing}, {Ross}, {Sand}, {Terndrup},
  {Valenti}, {Walker}, \& {Yoon}}]{shappee14}
{Shappee} B.~J. {et~al.}, 2014, \apj, 788, 48

\bibitem[{{Silverman} {et~al}\mbox{.}(2012){Silverman}, {Foley}, {Filippenko},
  {Ganeshalingam}, {Barth}, {Chornock}, {Griffith}, {Kong}, {Lee}, {Leonard},
  {Matheson}, {Miller}, {Steele}, {Barris}, {Bloom}, {Cobb}, {Coil},
  {Desroches}, {Gates}, {Ho}, {Jha}, {Kandrashoff}, {Li}, {Mandel}, {Modjaz},
  {Moore}, {Mostardi}, {Papenkova}, {Park}, {Perley}, {Poznanski}, {Reuter},
  {Scala}, {Serduke}, {Shields}, {Swift}, {Tonry}, {Van Dyk}, {Wang}, \&
  {Wong}}]{silverman12}
{Silverman} J.~M. {et~al.}, 2012, \mnras, 425, 1789

\bibitem[{{Smith} {et~al}\mbox{.}(2006){Smith}, {Nordsieck}, {Burgh},
  {Percival}, {Williams}, {O'Donohue}, {O'Connor}, \& {Schier}}]{smith06}
{Smith} M.~P., {Nordsieck} K.~H., {Burgh} E.~B., {Percival} J.~W., {Williams}
  T.~B., {O'Donohue} D., {O'Connor} J., {Schier} J.~A., 2006, in Society of
  Photo-Optical Instrumentation Engineers (SPIE) Conference Series, Vol. 6269,
  p. 62692A

\bibitem[{{Sorokina} {et~al}\mbox{.}(2016){Sorokina}, {Blinnikov}, {Nomoto},
  {Quimby}, \& {Tolstov}}]{sokorina16}
{Sorokina} E., {Blinnikov} S., {Nomoto} K., {Quimby} R., {Tolstov} A., 2016,
  \apj, 829, 17

\bibitem[{{Stoll} {et~al}\mbox{.}(2011){Stoll}, {Prieto}, {Stanek}, {Pogge},
  {Szczygie{\l}}, {Pojma{\'n}ski}, {Antognini}, \& {Yan}}]{stoll11}
{Stoll} R., {Prieto} J.~L., {Stanek} K.~Z., {Pogge} R.~W., {Szczygie{\l}}
  D.~M., {Pojma{\'n}ski} G., {Antognini} J., {Yan} H., 2011, \apj, 730, 34

\bibitem[{{Sukhbold} \& {Woosley}(2016)}]{sukhbold16}
{Sukhbold} T., {Woosley} S.~E., 2016, \apjl, 820, L38

\bibitem[{{van Velzen} \& {Farrar}(2014)}]{vanvelzen14}
{van Velzen} S., {Farrar} G.~R., 2014, \apj, 792, 53

\bibitem[{{Wade} \& {Horne}(1988)}]{wade88}
{Wade} R.~A., {Horne} K., 1988, \apj, 324, 411

\bibitem[{{Wang} {et~al}\mbox{.}(2016){Wang}, {Liu}, {Dai}, {Wang}, \&
  {Wu}}]{wang16}
{Wang} S.~Q., {Liu} L.~D., {Dai} Z.~G., {Wang} L.~J., {Wu} X.~F., 2016, \apj,
  828, 87

\bibitem[{{Woosley}(2010)}]{woosley10}
{Woosley} S.~E., 2010, \apjl, 719, L204

\bibitem[{{Yan} {et~al}\mbox{.}(2015){Yan}, {Quimby}, {Ofek}, {Gal-Yam},
  {Mazzali}, {Perley}, {Vreeswijk}, {Leloudas}, {de Cia}, {Masci}, {Cenko},
  {Cao}, {Kulkarni}, {Nugent}, {Rebbapragada}, {Woźniak}, \& {Yaron}}]{yan15}
{Yan} L. {et~al.}, 2015, \apj, 814, 108

\bibitem[{{Young} {et~al}\mbox{.}(2010){Young}, {Smartt}, {Valenti},
  {Pastorello}, {Benetti}, {Benn}, {Bersier}, {Botticella}, {Corradi},
  {Harutyunyan}, {Hrudkova}, {Hunter}, {Mattila}, {de Mooij}, {Navasardyan},
  {Snellen}, {Tanvir}, \& {Zampieri}}]{young10}
{Young} D.~R. {et~al.}, 2010, \aap, 512, A70

\end{thebibliography}

\newpage

\begin{table*}
\begin{minipage}{\textwidth}
\centering
\caption{{\swift} photometric data of {\name}.\hfill}
\renewcommand{\arraystretch}{1.2}
\begin{tabular}{ccccccccccccc}
\hline
MJD & $UVW2$ &  $\sigma_{UVW2}$ & $UVM2$ &  $\sigma_{UVM2}$ & $UVW1$ &  $\sigma_{UVW1}$ & $U$ &  $\sigma_{U}$ & $B$ &  $\sigma_{B}$ & $V$ &  $\sigma_{V}$ \\
\hline
57197.0 & 15.46 & 0.05 & 15.15 & 0.06 & 15.20 & 0.05 & 15.38 & 0.04 & 16.76 & 0.06 & 17.05 & 0.14 \\
57199.8 & 15.55 & 0.05 & 15.22 & 0.05 & 15.22 & 0.05 & 15.44 & 0.04 & 16.75 & 0.05 & 16.76 & 0.10 \\
57201.8 & 15.83 & 0.05 & 15.48 & 0.06 & 15.32 & 0.05 & 15.51 & 0.05 & 16.89 & 0.06 & 16.96 & 0.13 \\
57205.5 & 15.76 & 0.05 & 15.51 & 0.05 & 15.47 & 0.05 & 15.59 & 0.05 & 16.97 & 0.06 & 16.92 & 0.11 \\
57208.6 & 15.95 & 0.07 & 15.69 & 0.06 & 15.58 & 0.07 & 15.83 & 0.07 & 17.04 & 0.09 & 16.89 & 0.17 \\
57211.5 & 16.20 & 0.07 & 15.87 & 0.06 & 15.81 & 0.07 & 15.74 & 0.07 & 17.10 & 0.09 & 17.05 & 0.16 \\
57214.6 & 16.18 & 0.05 & 15.94 & 0.07 & 15.81 & 0.06 & 15.86 & 0.05 & 17.14 & 0.06 & 17.12 & 0.12 \\
57214.7 & 16.28 & 0.07 & 16.04 & 0.07 & 15.89 & 0.07 & 15.89 & 0.07 & 17.03 & 0.08 & 17.31 & 0.19 \\
57216.4 & 16.42 & 0.09 & - & - & - & - & 15.91 & 0.04 & - & - & - & - \\
57216.5 & - & - & - & - & - & - & 15.87 & 0.04 & - & - & - & - \\
57216.8 & - & - & - & - & - & - & 15.90 & 0.04 & - & - & - & - \\
57217.1 & 16.30 & 0.05 & - & - & - & - & - & - & - & - & - & - \\
57217.6 & 16.30 & 0.07 & - & - & - & - & - & - & - & - & - & - \\
57219.7 & 16.47 & 0.10 & - & - & 16.18 & 0.06 & - & - & - & - & - & - \\
57220.5 & 16.63 & 0.09 & - & - & - & - & 16.08 & 0.04 & - & - & - & - \\
57221.7 & 16.57 & 0.06 & 16.42 & 0.06 & 16.22 & 0.06 & 16.03 & 0.05 & 17.29 & 0.07 & 17.28 & 0.13 \\
57223.5 & 16.73 & 0.06 & 16.60 & 0.06 & 16.32 & 0.06 & 16.17 & 0.06 & 17.37 & 0.07 & 17.24 & 0.13 \\
57226.1 & 17.00 & 0.09 & 16.71 & 0.10 & 16.46 & 0.09 & 16.28 & 0.07 & 17.39 & 0.09 & 17.12 & 0.15 \\
57229.1 & 17.24 & 0.11 & 16.82 & 0.10 & 16.58 & 0.10 & 16.61 & 0.12 & 17.34 & 0.12 & 17.50 & 0.26 \\
57232.8 & 17.34 & 0.09 & 17.17 & 0.09 & 16.84 & 0.09 & 16.42 & 0.08 & 17.51 & 0.10 & 17.53 & 0.20 \\
57241.1 & 17.70 & 0.09 & 17.48 & 0.08 & 17.13 & 0.08 & 16.66 & 0.07 & 17.56 & 0.08 & 17.65 & 0.16 \\
57244.9 & 17.80 & 0.10 & 17.63 & 0.09 & 17.40 & 0.10 & 16.77 & 0.08 & 17.72 & 0.09 & 17.50 & 0.16 \\
57248.4 & 17.96 & 0.10 & 17.67 & 0.10 & 17.38 & 0.10 & 16.76 & 0.09 & 17.60 & 0.10 & 17.66 & 0.20 \\
57250.7 & 17.93 & 0.10 & 17.71 & 0.10 & 17.22 & 0.10 & 16.82 & 0.09 & 17.51 & 0.09 & 17.34 & 0.16 \\
57253.6 & 17.96 & 0.13 & 17.67 & 0.12 & 17.27 & 0.13 & 16.62 & 0.10 & 17.69 & 0.12 & 17.37 & 0.21 \\
57255.6 & 18.08 & 0.10 & 17.89 & 0.13 & 17.41 & 0.10 & 16.88 & 0.07 & 17.77 & 0.08 & 17.87 & 0.18 \\
57259.3 & 18.43 & 0.16 & 17.94 & 0.13 & 17.43 & 0.13 & 17.05 & 0.12 & 17.95 & 0.14 & 17.78 & 0.27 \\
57262.6 & 18.19 & 0.16 & 18.01 & 0.14 & 17.38 & 0.13 & 17.03 & 0.13 & 17.83 & 0.14 & 17.57 & 0.25 \\
57265.5 & 18.76 & 0.33 & - & - & 17.42 & 0.10 & 16.99 & 0.09 & 17.97 & 0.11 & - & - \\
57268.7 & 18.30 & 0.12 & 17.96 & 0.10 & 17.52 & 0.10 & 17.02 & 0.09 & 17.83 & 0.10 & 17.55 & 0.17 \\
57272.4 & 18.23 & 0.12 & 17.78 & 0.13 & 17.26 & 0.10 & 16.88 & 0.08 & 17.87 & 0.10 & 17.57 & 0.16 \\
57272.8 & 18.00 & 0.10 & 17.88 & 0.16 & 17.43 & 0.10 & 16.99 & 0.09 & 17.85 & 0.11 & 17.62 & 0.19 \\
57277.3 & 17.70 & 0.08 & 17.57 & 0.10 & 17.05 & 0.08 & 16.88 & 0.06 & 17.74 & 0.07 & 17.51 & 0.11 \\
57282.8 & 17.60 & 0.10 & 17.51 & 0.13 & 17.09 & 0.10 & 16.91 & 0.09 & - & - & - & - \\
57283.1 & 17.79 & 0.10 & 17.14 & 0.11 & 17.18 & 0.10 & 17.01 & 0.09 & - & - & - & - \\
57284.2 & 17.60 & 0.10 & 17.26 & 0.09 & 17.09 & 0.10 & 16.98 & 0.10 & 17.85 & 0.12 & 17.91 & 0.25 \\
57284.6 & 17.66 & 0.10 & 17.29 & 0.12 & 17.04 & 0.10 & 16.85 & 0.09 & 17.89 & 0.11 & 18.13 & 0.26 \\
57290.8 & 17.60 & 0.09 & 17.23 & 0.10 & 17.06 & 0.09 & 16.87 & 0.07 & - & - & - & - \\
57293.5 & 17.46 & 0.07 & 16.95 & 0.07 & 16.92 & 0.07 & 16.99 & 0.06 & - & - & - & - \\
57296.5 & 17.39 & 0.08 & 16.87 & 0.08 & 16.78 & 0.08 & 17.00 & 0.07 & - & - & - & - \\
57298.6 & 17.27 & 0.07 & 16.96 & 0.06 & 16.89 & 0.08 & 16.94 & 0.07 & 18.10 & 0.10 & 17.73 & 0.16 \\
57299.4 & 17.34 & 0.08 & 16.89 & 0.09 & 16.80 & 0.09 & 16.90 & 0.08 & - & - & - & - \\
57302.4 & 17.46 & 0.10 & 17.15 & 0.13 & 17.01 & 0.11 & 17.07 & 0.11 & - & - & - & - \\
57305.4 & 17.13 & 0.06 & 17.04 & 0.08 & 16.77 & 0.07 & 16.88 & 0.06 & 17.97 & 0.08 & 17.77 & 0.14 \\
57307.4 & 17.15 & 0.07 & 16.84 & 0.06 & 16.82 & 0.07 & 16.92 & 0.08 & 18.06 & 0.10 & 18.07 & 0.21 \\
57309.7 & 17.17 & 0.07 & 16.85 & 0.06 & 16.72 & 0.08 & 16.84 & 0.08 & 18.29 & 0.13 & 17.52 & 0.15 \\
57311.3 & 17.01 & 0.07 & 16.75 & 0.06 & 16.78 & 0.08 & 17.04 & 0.09 & 18.17 & 0.12 & 17.66 & 0.17 \\
57313.1 & 17.13 & 0.10 & 16.81 & 0.08 & 16.74 & 0.10 & 16.96 & 0.12 & 18.11 & 0.16 & 17.94 & 0.30 \\
57316.2 & 17.13 & 0.09 & 16.94 & 0.11 & 16.70 & 0.10 & 16.77 & 0.09 & - & - & - & - \\
57319.7 & 17.26 & 0.07 & 17.02 & 0.09 & 16.88 & 0.08 & 16.95 & 0.07 & - & - & - & - \\
57322.7 & - & - & 17.02 & 0.09 & 16.71 & 0.12 & - & - & - & - & - & - \\
57325.3 & 17.39 & 0.06 & 17.11 & 0.08 & 16.89 & 0.07 & 16.94 & 0.06 & - & - & - & - \\
57328.3 & 17.24 & 0.10 & 16.79 & 0.11 & 16.84 & 0.11 & 16.74 & 0.11 & - & - & - & - \\
57331.0 & 17.19 & 0.08 & 16.87 & 0.09 & 16.84 & 0.09 & 16.88 & 0.08 & - & - & - & - \\
57334.5 & 17.12 & 0.07 & 16.86 & 0.09 & 16.75 & 0.08 & 16.94 & 0.08 & - & - & - & - \\
\hline
\end{tabular}
\label{tab:table1}
\end{minipage}
\end{table*}

\begin{table*}
\begin{minipage}{\textwidth}
\centering
\renewcommand{\arraystretch}{1.2}
\begin{tabular}{ccccccccccccc}
\hline
MJD & $UVW2$ &  $\sigma_{UVW2}$ & $UVM2$ &  $\sigma_{UVM2}$ & $UVW1$ &  $\sigma_{UVW1}$ & $U$ &  $\sigma_{U}$ & $B$ &  $\sigma_{B}$ & $V$ &  $\sigma_{V}$ \\
\hline
57343.4 & 17.18 & 0.07 & 16.94 & 0.08 & 16.85 & 0.07 & 17.00 & 0.06 & - & - & - & - \\
57349.2 & 17.10 & 0.07 & 16.82 & 0.08 & 16.79 & 0.08 & 17.06 & 0.07 & - & - & - & - \\
57353.5 & 17.16 & 0.10 & 16.86 & 0.10 & 16.85 & 0.09 & 16.93 & 0.13 & - & - & - & - \\
57356.2 & 17.03 & 0.06 & - & - & 16.92 & 0.08 & 16.96 & 0.07 & - & - & - & - \\
57359.3 & 17.14 & 0.07 & 16.98 & 0.09 & 16.90 & 0.08 & 16.86 & 0.07 & - & - & - & - \\
57362.2 & 17.00 & 0.12 & 16.95 & 0.15 & 16.73 & 0.13 & 17.20 & 0.18 & - & - & - & - \\
57366.4 & 17.12 & 0.17 & - & - & 17.06 & 0.21 & 17.15 & 0.11 & 18.31 & 0.19 & - & - \\
57368.9 & 17.16 & 0.08 & 17.31 & 0.11 & 16.87 & 0.09 & 17.03 & 0.09 & - & - & - & - \\
57371.4 & 17.28 & 0.07 & 17.06 & 0.10 & 17.15 & 0.09 & 17.21 & 0.08 & 18.40 & 0.13 & - & - \\
57374.4 & 17.36 & 0.07 & 16.98 & 0.09 & 17.11 & 0.09 & 17.22 & 0.09 & - & - & - & - \\
57377.2 & 17.30 & 0.06 & 17.11 & 0.08 & 17.29 & 0.10 & 17.30 & 0.09 & - & - & - & - \\
57380.9 & 17.36 & 0.09 & 16.98 & 0.10 & 17.12 & 0.10 & 17.24 & 0.12 & - & - & - & - \\
57383.2 & 17.42 & 0.10 & 17.43 & 0.19 & 17.18 & 0.12 & 17.29 & 0.15 & - & - & - & - \\
57385.9 & 17.31 & 0.11 & 17.14 & 0.15 & 17.14 & 0.15 & 17.19 & 0.19 & - & - & - & - \\
57393.3 & 17.35 & 0.11 & 17.21 & 0.19 & 17.40 & 0.17 & 17.09 & 0.15 & - & - & - & - \\
57396.7 & 17.67 & 0.08 & 17.34 & 0.09 & 17.36 & 0.10 & 17.34 & 0.10 & 18.46 & 0.16 & 18.14 & 0.22 \\
57399.2 & 17.47 & 0.08 & 17.47 & 0.10 & 17.42 & 0.10 & 17.34 & 0.11 & - & - & 18.08 & 0.25 \\
57427.7 & 17.95 & 0.11 & 17.70 & 0.12 & 17.63 & 0.14 & 17.74 & 0.17 & 18.69 & 0.26 & - & - \\
57429.4 & 17.90 & 0.15 & 17.53 & 0.15 & 17.82 & 0.22 & 17.99 & 0.29 & 18.26 & 0.26 & - & - \\
57433.1 & 18.27 & 0.19 & 18.07 & 0.20 & 17.59 & 0.19 & 17.90 & 0.31 & - & - & - & - \\
57437.1 & 17.92 & 0.10 & 17.97 & 0.24 & 17.56 & 0.12 & 17.68 & 0.15 & 18.53 & 0.23 & - & - \\
57447.2 & 18.04 & 0.13 & 17.96 & 0.16 & 17.74 & 0.17 & 17.75 & 0.21 & - & - & - & - \\
57449.9 & 18.18 & 0.12 & 17.94 & 0.14 & 17.95 & 0.16 & 17.73 & 0.19 & - & - & - & - \\
57450.2 & 18.20 & 0.12 & 18.14 & 0.16 & 18.03 & 0.17 & 17.65 & 0.17 & 18.67 & 0.27 & - & - \\
57453.9 & 17.92 & 0.11 & 17.91 & 0.14 & 17.64 & 0.14 & 17.71 & 0.18 & 18.57 & 0.27 & - & - \\
57456.0 & 17.96 & 0.10 & 17.88 & 0.14 & 17.81 & 0.13 & 18.05 & 0.18 & 18.72 & 0.22 & - & - \\
57457.5 & 18.37 & 0.18 & - & - & 18.10 & 0.21 & 17.65 & 0.18 & 18.79 & 0.31 & - & - \\
57465.1 & 18.16 & 0.14 & 18.17 & 0.20 & 17.94 & 0.18 & 17.83 & 0.19 & - & - & - & - \\
57470.2 & 18.15 & 0.10 & 17.89 & 0.11 & 17.89 & 0.11 & 17.69 & 0.10 & - & - & - & - \\
57472.2 & 18.19 & 0.13 & 18.04 & 0.16 & 17.84 & 0.15 & 17.76 & 0.15 & 19.02 & 0.27 & 18.18 & 0.33 \\
57475.6 & 18.21 & 0.11 & 17.90 & 0.13 & 17.88 & 0.13 & 17.86 & 0.13 & - & - & - & - \\
57477.2 & 18.37 & 0.18 & 17.91 & 0.19 & 18.03 & 0.22 & 17.68 & 0.21 & 18.59 & 0.29 & - & - \\
57480.0 & 18.45 & 0.14 & 18.16 & 0.16 & 18.12 & 0.17 & 18.06 & 0.21 & - & - & - & - \\
57487.1 & 18.44 & 0.18 & 18.33 & 0.23 & 17.84 & 0.18 & 18.00 & 0.25 & - & - & - & - \\
57489.0 & 18.30 & 0.10 & 18.26 & 0.13 & 18.27 & 0.13 & 18.00 & 0.13 & - & - & - & - \\
57499.7 & - & - & 18.48 & 0.16 & 18.35 & 0.27 & - & - & - & - & - & - \\
57502.7 & 18.42 & 0.13 & 18.21 & 0.16 & 18.02 & 0.16 & 18.03 & 0.20 & - & - & - & - \\
57505.2 & 18.45 & 0.14 & 18.39 & 0.19 & 17.89 & 0.15 & 18.00 & 0.18 & - & - & - & - \\
57508.9 & 18.41 & 0.11 & 18.27 & 0.14 & 17.92 & 0.12 & 17.77 & 0.12 & - & - & - & - \\
57511.8 & 18.49 & 0.11 & 18.46 & 0.14 & 18.12 & 0.12 & 18.19 & 0.15 & - & - & - & - \\
57515.9 & 18.51 & 0.11 & 18.37 & 0.14 & 18.07 & 0.11 & 18.05 & 0.11 & - & - & - & - \\
57516.3 & 18.38 & 0.15 & 18.20 & 0.18 & 18.24 & 0.19 & 17.82 & 0.18 & - & - & - & - \\
57520.3 & 18.70 & 0.19 & 18.38 & 0.21 & 18.15 & 0.19 & 18.42 & 0.30 & - & - & - & - \\
57523.7 & 18.42 & 0.13 & 18.45 & 0.19 & 18.27 & 0.17 & 18.24 & 0.16 & - & - & - & - \\
57527.9 & 18.40 & 0.13 & 18.33 & 0.17 & 18.14 & 0.16 & 18.52 & 0.25 & - & - & - & - \\
57528.9 & 18.92 & 0.21 & 18.71 & 0.25 & 18.38 & 0.22 & 18.22 & 0.24 & - & - & - & - \\
57533.6 & 18.72 & 0.25 & 19.17 & 0.30 & 18.09 & 0.22 & 18.43 & 0.33 & 18.77 & 0.30 & - & - \\
57552.5 & 18.58 & 0.18 & 18.35 & 0.15 & 18.89 & 0.27 & 18.58 & 0.28 & 18.95 & 0.26 & - & - \\
57554.8 & 18.64 & 0.18 & 18.44 & 0.16 & 18.62 & 0.23 & 18.25 & 0.22 & - & - & 18.01 & 0.29 \\
57557.2 & 18.86 & 0.18 & 18.56 & 0.15 & 18.61 & 0.20 & 18.40 & 0.22 & 19.38 & 0.31 & - & - \\
57560.0 & 18.67 & 0.13 & 18.51 & 0.13 & 18.44 & 0.14 & 18.24 & 0.16 & 19.33 & 0.24 & 18.63 & 0.32 \\
57563.0 & 18.63 & 0.18 & 18.39 & 0.15 & 18.59 & 0.24 & 18.15 & 0.22 & 19.15 & 0.31 & - & - \\
57569.1 & 18.89 & 0.23 & 18.84 & 0.22 & 18.11 & 0.18 & 18.17 & 0.23 & - & - & 18.10 & 0.35 \\
57577.8 & 18.86 & 0.19 & 18.53 & 0.22 & 18.12 & 0.17 & 18.45 & 0.21 & - & - & - & - \\
57591.7 & 19.29 & 0.22 & 19.01 & 0.27 & 18.94 & 0.26 & 18.40 & 0.19 & - & - & - & - \\
57600.4 & 18.90 & 0.17 & 19.05 & 0.24 & 18.87 & 0.22 & 18.50 & 0.18 & - & - & - & - \\
\hline
\end{tabular}
\end{minipage}
\end{table*}

\begin{table*}
\begin{minipage}{\textwidth}
\centering
\renewcommand{\arraystretch}{1.2}
\begin{tabular}{ccccccccccccc}
\hline
MJD & $UVW2$ &  $\sigma_{UVW2}$ & $UVM2$ &  $\sigma_{UVM2}$ & $UVW1$ &  $\sigma_{UVW1}$ & $U$ &  $\sigma_{U}$ & $B$ &  $\sigma_{B}$ & $V$ &  $\sigma_{V}$ \\
\hline
57601.4 & 19.26 & 0.21 & 19.13 & 0.25 & 18.99 & 0.23 & 18.53 & 0.18 & - & - & - & - \\
57608.0 & 19.33 & 0.32 & 18.73 & 0.25 & 18.60 & 0.24 & 18.85 & 0.33 & - & - & - & - \\
57609.0 & 19.01 & 0.16 & 19.16 & 0.23 & 18.59 & 0.16 & 18.57 & 0.17 & - & - & - & - \\
57611.5 & 19.27 & 0.17 & 18.90 & 0.19 & 19.13 & 0.22 & 18.86 & 0.20 & - & - & - & - \\
57615.4 & 18.94 & 0.12 & 18.87 & 0.16 & 18.72 & 0.15 & 18.79 & 0.16 & - & - & - & - \\
57622.5 & 19.16 & 0.26 & - & - & 18.76 & 0.20 & 18.70 & 0.15 & - & - & - & - \\
57626.7 & 19.38 & 0.23 & 19.42 & 0.36 & 18.96 & 0.28 & - & - & - & - & - & - \\
57630.4 & 19.03 & 0.23 & - & - & - & - & 18.50 & 0.26 & - & - & - & - \\
57632.8 & 18.92 & 0.20 & - & - & - & - & 18.86 & 0.28 & - & - & - & - \\
57640.0 & 19.58 & 0.26 & 19.13 & 0.26 & 18.89 & 0.25 & 18.65 & 0.20 & - & - & - & - \\
57645.5 & 19.56 & 0.24 & 19.36 & 0.32 & 18.76 & 0.23 & 18.78 & 0.24 & - & - & - & - \\
57654.2 & 19.35 & 0.32 & - & - & 18.64 & 0.31 & 18.73 & 0.34 & - & - & - & - \\
57667.3 & 19.31 & 0.31 & - & - & - & - & - & - & - & - & - & - \\
57677.3 & 19.37 & 0.32 & - & - & - & - & - & - & - & - & - & - \\
57680.2 & 19.21 & 0.30 & - & - & 18.72 & 0.36 & - & - & - & - & - & - \\
57682.5 & 19.32 & 0.15 & 19.36 & 0.23 & 19.13 & 0.20 & 19.11 & 0.21 & - & - & - & - \\
57687.0 & 19.19 & 0.24 & - & - & - & - & - & - & - & - & - & - \\
57688.9 & 19.31 & 0.25 & 18.92 & 0.31 & - & - & 19.06 & 0.36 & - & - & - & - \\
57704.0 & 19.93 & 0.33 & - & - & - & - & 18.85 & 0.27 & - & - & - & - \\
\hline
\end{tabular}
\end{minipage}
\end{table*}

\begin{table*}
\begin{minipage}{\textwidth}
\centering
\caption{LCOGT photometric data of {\name.}\hfill}
\renewcommand{\arraystretch}{1.2}
\begin{tabular}{ccccc}
\hline
MJD & $B$ &  $\sigma_{B}$ & $V$ &  $\sigma_{V}$ \\
\hline
57190.4 & 16.78 & 0.07 & 16.84 & 0.04 \\
57196.1 & 16.75 & 0.03 & 16.85 & 0.03 \\
57196.2 & 16.72 & 0.03 & 16.90 & 0.03 \\
57200.8 & 16.86 & 0.05 & 16.91 & 0.04 \\
57203.1 & 16.83 & 0.05 & 16.97 & 0.04 \\
57204.2 & 16.86 & 0.04 & 16.96 & 0.04 \\
57206.2 & 16.88 & 0.05 & 17.02 & 0.05 \\
57208.6 & 16.92 & 0.08 & 17.06 & 0.04 \\
57210.3 & 16.95 & 0.05 & - & - \\
57212.5 & 17.07 & 0.04 & 17.13 & 0.04 \\
57216.9 & 17.06 & 0.05 & 17.14 & 0.06 \\
57219.0 & 17.11 & 0.03 & 17.24 & 0.04 \\
57221.2 & 17.29 & 0.04 & 17.30 & 0.03 \\
57222.9 & 17.25 & 0.04 & 17.35 & 0.05 \\
57225.4 & 17.38 & 0.07 & - & - \\
57227.3 & 17.34 & 0.07 & - & - \\
57228.7 & 17.33 & 0.06 & - & - \\
57230.5 & 17.36 & 0.07 & - & - \\
57234.5 & 17.50 & 0.08 & - & - \\
57240.0 & 17.59 & 0.05 & 17.48 & 0.04 \\
57241.0 & 17.51 & 0.04 & 17.48 & 0.04 \\
57242.5 & 17.58 & 0.05 & 17.49 & 0.04 \\
57244.5 & - & - & 17.55 & 0.04 \\
57245.4 & 17.60 & 0.04 & 17.55 & 0.04 \\
57247.5 & 17.72 & 0.04 & 17.52 & 0.04 \\
57250.5 & 17.63 & 0.04 & 17.61 & 0.04 \\
57253.5 & 17.66 & 0.03 & 17.56 & 0.04 \\
57256.6 & - & - & 17.65 & 0.05 \\
57258.0 & 17.79 & 0.08 & - & - \\
57261.1 & 17.79 & 0.05 & 17.66 & 0.05 \\
57262.3 & 17.99 & 0.09 & - & - \\
57262.6 & 17.88 & 0.07 & 17.66 & 0.07 \\
57264.1 & 17.99 & 0.08 & 17.65 & 0.06 \\
57267.2 & 17.90 & 0.05 & 17.68 & 0.04 \\
57268.1 & 17.87 & 0.05 & 17.73 & 0.03 \\
57270.2 & 17.90 & 0.05 & - & - \\
57274.0 & - & - & 17.70 & 0.04 \\
57279.0 & 17.89 & 0.07 & 17.66 & 0.05 \\
57282.3 & 17.93 & 0.05 & 17.75 & 0.03 \\
57285.0 & - & - & 17.88 & 0.06 \\
57286.0 & 17.95 & 0.04 & 17.76 & 0.04 \\
57288.3 & 17.90 & 0.05 & 17.68 & 0.04 \\
57288.6 & 17.85 & 0.05 & 17.78 & 0.04 \\
57291.1 & 17.96 & 0.06 & 17.82 & 0.05 \\
57296.5 & 18.01 & 0.05 & 17.79 & 0.04 \\
57298.6 & 17.92 & 0.08 & 17.84 & 0.04 \\
57301.5 & - & - & 17.81 & 0.04 \\
57302.9 & 17.97 & 0.04 & 17.79 & 0.04 \\
57304.5 & 17.98 & 0.04 & - & - \\
57305.5 & 17.92 & 0.04 & 17.81 & 0.03 \\
57307.9 & 18.01 & 0.05 & 17.83 & 0.04 \\
57311.0 & 17.96 & 0.04 & 17.81 & 0.04 \\
57319.8 & 17.90 & 0.06 & 17.92 & 0.06 \\
57328.8 & 18.01 & 0.05 & 17.87 & 0.04 \\
57332.1 & 18.03 & 0.05 & 17.86 & 0.03 \\
\hline
\end{tabular}
\label{tab:table2}
\end{minipage}
\end{table*}

\begin{table*}
\begin{minipage}{\textwidth}
\centering
\renewcommand{\arraystretch}{1.2}
\begin{tabular}{ccccc}
\hline
MJD & $B$ &  $\sigma_{B}$ & $V$ &  $\sigma_{V}$ \\
\hline
57334.8 & 18.04 & 0.05 & 17.87 & 0.03 \\
57338.1 & 18.13 & 0.05 & 17.88 & 0.03 \\
57345.5 & 18.14 & 0.06 & 18.02 & 0.05 \\
57350.8 & 18.06 & 0.06 & 17.94 & 0.05 \\
57353.8 & 18.12 & 0.05 & 17.95 & 0.04 \\
57356.8 & 18.12 & 0.04 & 17.97 & 0.04 \\
57360.5 & 18.27 & 0.04 & 18.04 & 0.03 \\
57363.0 & 18.17 & 0.08 & 18.02 & 0.03 \\
57366.8 & 18.23 & 0.04 & 18.05 & 0.04 \\
57373.1 & 18.27 & 0.05 & 18.06 & 0.03 \\
57375.4 & 18.29 & 0.05 & 18.08 & 0.04 \\
57378.0 & - & - & 18.10 & 0.04 \\
57379.0 & - & - & 18.13 & 0.04 \\

57492.1 & - & - & 18.50 & 0.05 \\
57563.4 & 19.03 & 0.08 & 18.61 & 0.06 \\
57581.2 & - & - & 18.63 & 0.06 \\
57587.1 & 19.28 & 0.10 & 18.59 & 0.06 \\
57598.4 & - & - & 18.69 & 0.05 \\
57598.7 & 19.37 & 0.09 & 18.75 & 0.06 \\
57598.8 & 19.44 & 0.09 & - & - \\
57602.1 & - & - & 18.53 & 0.05 \\
57607.2 & 19.58 & 0.07 & 18.69 & 0.11 \\
57612.8 & - & - & 18.71 & 0.06 \\
57613.7 & 19.46 & 0.08 & - & - \\
57624.0 & 19.52 & 0.11 & 18.75 & 0.06 \\
57634.2 & - & - & 18.77 & 0.05 \\
57637.9 & 19.60 & 0.09 & 18.76 & 0.05 \\
57643.3 & 19.56 & 0.09 & 18.84 & 0.05 \\
57654.3 & - & - & 18.74 & 0.06 \\
57663.2 & - & - & 18.80 & 0.05 \\
57668.1 & 19.51 & 0.08 & 18.75 & 0.06 \\
57670.2 & 19.51 & 0.09 & 18.83 & 0.06 \\
57672.8 & 19.44 & 0.20 & - & - \\
57680.1 & - & - & 18.73 & 0.06 \\
57687.6 & - & - & 18.77 & 0.06 \\
57687.9 & 19.61 & 0.07 & 18.77 & 0.05 \\
57694.5 & 19.76 & 0.07 & 18.85 & 0.05 \\
57699.9 & 19.57 & 0.08 & - & - \\
\hline
\end{tabular}
\end{minipage}
\end{table*}

\begin{table*}
\begin{minipage}{\textwidth}
\centering
\caption{Blackbody models for the bolometric luminosity, effective temperature, and apparent photospheric radius of {\name} using the constant-temperature prior for early-times ($t<$12.5 days). The units of the variables are solar luminosities, Kelvin, and centimeters, respectively. The logarithms are all in base 10.\hfill}
\renewcommand{\arraystretch}{1.2}
\begin{tabular}{ccccccccccc}
\hline
MJD & $N_{\text{obs}}$ & $\log(L)$ & $\log(L_{-})$  &  $\log(L_{+})$ & $\log(T)$  & $\log(T_{-})$  & $\log(T_{+})$ & $\log(R)$  & $\log(R_{-})$  & $\log(R_{+})$ \\
\hline
57150.5 & 1 & 11.43 & 11.26 & 11.62 & 4.32 & 4.24 & 4.40 & 15.45 & 15.36 & 15.54 \\
57157.0 & 1 & 11.27 & 11.09 & 11.45 & 4.32 & 4.24 & 4.40 & 15.36 & 15.27 & 15.45 \\
57161.1 & 1 & 11.53 & 11.36 & 11.73 & 4.32 & 4.24 & 4.40 & 15.50 & 15.41 & 15.59 \\
57169.8 & 1 & 11.75 & 11.58 & 11.93 & 4.32 & 4.24 & 4.40 & 15.60 & 15.52 & 15.69 \\
57172.7 & 1 & 11.64 & 11.47 & 11.83 & 4.32 & 4.24 & 4.40 & 15.55 & 15.46 & 15.64 \\
57181.8 & 1 & 11.75 & 11.58 & 11.94 & 4.32 & 4.24 & 4.40 & 15.60 & 15.52 & 15.70 \\
57184.0 & 1 & 11.66 & 11.49 & 11.85 & 4.32 & 4.24 & 4.40 & 15.56 & 15.47 & 15.65 \\
57190.7 & 2 & 11.61 & 11.44 & 11.79 & 4.32 & 4.24 & 4.40 & 15.53 & 15.45 & 15.62 \\
\hline
\end{tabular}
\label{tab:table3}
\end{minipage}
\end{table*}

\begin{table*}
\begin{minipage}{\textwidth}
\centering
\caption{Blackbody models for the bolometric luminosity, effective temperature, and apparent photospheric radius of {\name} using the rising-temperature prior for early-times ($t<$12.5 days). The units of the variables are solar luminosities, Kelvin, and centimeters, respectively. The logarithms are all in base 10.\hfill}
\renewcommand{\arraystretch}{1.2}
\begin{tabular}{ccccccccccc}
\hline
MJD & $N_{\text{obs}}$ & $\log(L)$ & $\log(L_{-})$  &  $\log(L_{+})$ & $\log(T)$  & $\log(T_{-})$  & $\log(T_{+})$ & $\log(R)$  & $\log(R_{-})$  & $\log(R_{+})$ \\
\hline
57150.5 & 1 & 11.69 & 11.50 & 11.89 & 4.44 & 4.36 & 4.52 & 15.34 & 15.26 & 15.42 \\
57157.0 & 1 & 11.49 & 11.30 & 11.69 & 4.42 & 4.34 & 4.51 & 15.27 & 15.19 & 15.35 \\
57161.1 & 1 & 11.73 & 11.54 & 11.93 & 4.41 & 4.33 & 4.49 & 15.41 & 15.33 & 15.50 \\
57169.8 & 1 & 11.89 & 11.71 & 12.09 & 4.39 & 4.31 & 4.47 & 15.54 & 15.46 & 15.62 \\
57172.7 & 1 & 11.77 & 11.58 & 11.96 & 4.38 & 4.30 & 4.46 & 15.49 & 15.41 & 15.58 \\
57181.8 & 1 & 11.81 & 11.64 & 12.00 & 4.35 & 4.27 & 4.43 & 15.57 & 15.49 & 15.66 \\
57184.0 & 1 & 11.71 & 11.53 & 11.90 & 4.34 & 4.26 & 4.43 & 15.54 & 15.45 & 15.63 \\
57190.7 & 2 & 11.61 & 11.44 & 11.79 & 4.32 & 4.24 & 4.41 & 15.53 & 15.45 & 15.62 \\
\hline
\end{tabular}
\label{tab:table4}
\end{minipage}
\end{table*}

\begin{table*}
\begin{minipage}{\textwidth}
\centering
\caption{Blackbody models for the bolometric luminosity, effective temperature, and apparent photospheric radius of {\name} for times after $t=$12.5 days, when no prior is required. The units of the variables are solar luminosities, Kelvin, and centimeters, respectively. The logarithms are all in base 10. \hfill}
\renewcommand{\arraystretch}{1.2}
\begin{tabular}{ccccccccccc}
\hline
MJD & $N_{\text{obs}}$ & $\log(L)$ & $\log(L_{-})$  &  $\log(L_{+})$ & $\log(T)$  & $\log(T_{-})$  & $\log(T_{+})$ & $\log(R)$  & $\log(R_{-})$  & $\log(R_{+})$ \\
\hline
57197.0 & 6 & 11.79 & 11.75 & 11.84 & 4.32 & 4.28 & 4.35 & 15.63 & 15.57 & 15.69 \\
57199.8 & 6 & 11.76 & 11.73 & 11.80 & 4.27 & 4.24 & 4.30 & 15.72 & 15.66 & 15.77 \\
57201.8 & 6 & 11.68 & 11.65 & 11.71 & 4.25 & 4.22 & 4.29 & 15.70 & 15.65 & 15.76 \\
57205.5 & 6 & 11.67 & 11.63 & 11.70 & 4.26 & 4.23 & 4.29 & 15.69 & 15.63 & 15.75 \\
57208.6 & 6 & 11.61 & 11.57 & 11.64 & 4.23 & 4.21 & 4.26 & 15.71 & 15.65 & 15.76 \\
57211.5 & 6 & 11.54 & 11.51 & 11.58 & 4.21 & 4.18 & 4.24 & 15.72 & 15.66 & 15.77 \\
57214.7 & 6 & 11.52 & 11.48 & 11.55 & 4.22 & 4.19 & 4.25 & 15.69 & 15.63 & 15.75 \\
57221.7 & 6 & 11.40 & 11.37 & 11.43 & 4.18 & 4.16 & 4.21 & 15.71 & 15.65 & 15.76 \\
57223.5 & 6 & 11.36 & 11.33 & 11.39 & 4.16 & 4.14 & 4.19 & 15.73 & 15.67 & 15.78 \\
57226.1 & 6 & 11.34 & 11.31 & 11.37 & 4.13 & 4.11 & 4.15 & 15.78 & 15.73 & 15.84 \\
57229.1 & 6 & 11.25 & 11.22 & 11.28 & 4.14 & 4.12 & 4.17 & 15.71 & 15.66 & 15.77 \\
57232.8 & 6 & 11.21 & 11.18 & 11.24 & 4.12 & 4.10 & 4.14 & 15.74 & 15.68 & 15.79 \\
57241.1 & 6 & 11.13 & 11.10 & 11.17 & 4.09 & 4.07 & 4.12 & 15.75 & 15.70 & 15.80 \\
57244.9 & 6 & 11.11 & 11.08 & 11.15 & 4.07 & 4.05 & 4.09 & 15.78 & 15.73 & 15.83 \\
57248.4 & 6 & 11.10 & 11.06 & 11.14 & 4.07 & 4.05 & 4.09 & 15.78 & 15.73 & 15.83 \\
57250.7 & 6 & 11.17 & 11.13 & 11.21 & 4.05 & 4.03 & 4.07 & 15.86 & 15.81 & 15.91 \\
57253.6 & 6 & 11.16 & 11.12 & 11.20 & 4.05 & 4.03 & 4.07 & 15.85 & 15.80 & 15.90 \\
57255.6 & 6 & 11.02 & 10.98 & 11.06 & 4.08 & 4.06 & 4.10 & 15.72 & 15.67 & 15.77 \\
57259.3 & 6 & 11.00 & 10.96 & 11.04 & 4.06 & 4.04 & 4.08 & 15.75 & 15.70 & 15.80 \\
57262.6 & 6 & 11.06 & 11.02 & 11.10 & 4.05 & 4.03 & 4.07 & 15.80 & 15.75 & 15.85 \\
57268.7 & 6 & 11.06 & 11.02 & 11.10 & 4.04 & 4.02 & 4.06 & 15.82 & 15.76 & 15.87 \\
57272.6 & 6 & 11.06 & 11.03 & 11.10 & 4.07 & 4.05 & 4.08 & 15.77 & 15.72 & 15.82 \\
57277.3 & 6 & 11.12 & 11.08 & 11.15 & 4.09 & 4.07 & 4.11 & 15.74 & 15.69 & 15.80 \\
57282.9 & 6 & 11.06 & 11.03 & 11.10 & 4.13 & 4.11 & 4.16 & 15.63 & 15.58 & 15.69 \\
57284.4 & 6 & 11.04 & 11.01 & 11.07 & 4.17 & 4.14 & 4.20 & 15.55 & 15.49 & 15.60 \\
57290.8 & 6 & 11.07 & 11.04 & 11.10 & 4.15 & 4.13 & 4.17 & 15.61 & 15.55 & 15.66 \\
57293.5 & 6 & 11.10 & 11.06 & 11.13 & 4.18 & 4.16 & 4.21 & 15.55 & 15.50 & 15.61 \\
57296.5 & 6 & 11.12 & 11.08 & 11.15 & 4.20 & 4.17 & 4.22 & 15.54 & 15.48 & 15.59 \\
57299.0 & 6 & 11.11 & 11.08 & 11.14 & 4.20 & 4.17 & 4.23 & 15.53 & 15.47 & 15.58 \\
57302.4 & 6 & 11.07 & 11.04 & 11.11 & 4.17 & 4.14 & 4.20 & 15.57 & 15.51 & 15.62 \\
57305.4 & 6 & 11.13 & 11.10 & 11.16 & 4.21 & 4.18 & 4.23 & 15.53 & 15.47 & 15.58 \\
57307.4 & 6 & 11.12 & 11.09 & 11.16 & 4.26 & 4.23 & 4.30 & 15.40 & 15.34 & 15.46 \\
57309.7 & 6 & 11.15 & 11.12 & 11.18 & 4.21 & 4.18 & 4.23 & 15.53 & 15.48 & 15.59 \\
57311.3 & 6 & 11.15 & 11.12 & 11.19 & 4.24 & 4.21 & 4.27 & 15.47 & 15.41 & 15.53 \\
57313.1 & 6 & 11.13 & 11.10 & 11.17 & 4.25 & 4.22 & 4.29 & 15.43 & 15.37 & 15.48 \\
57316.2 & 6 & 11.15 & 11.12 & 11.18 & 4.21 & 4.19 & 4.24 & 15.52 & 15.46 & 15.57 \\
57319.7 & 6 & 11.11 & 11.07 & 11.14 & 4.20 & 4.18 & 4.23 & 15.52 & 15.46 & 15.57 \\
57325.3 & 6 & 11.09 & 11.06 & 11.12 & 4.19 & 4.16 & 4.21 & 15.54 & 15.48 & 15.59 \\
57328.3 & 6 & 11.14 & 11.11 & 11.17 & 4.21 & 4.19 & 4.24 & 15.51 & 15.45 & 15.57 \\
57331.0 & 6 & 11.13 & 11.10 & 11.16 & 4.22 & 4.19 & 4.25 & 15.50 & 15.44 & 15.55 \\
57334.5 & 6 & 11.14 & 11.11 & 11.17 & 4.23 & 4.20 & 4.26 & 15.47 & 15.42 & 15.53 \\
57343.4 & 6 & 11.10 & 11.07 & 11.13 & 4.24 & 4.21 & 4.27 & 15.44 & 15.38 & 15.49 \\
57349.2 & 6 & 11.13 & 11.09 & 11.16 & 4.25 & 4.22 & 4.29 & 15.42 & 15.37 & 15.48 \\
57353.5 & 6 & 11.12 & 11.09 & 11.15 & 4.24 & 4.21 & 4.27 & 15.45 & 15.39 & 15.50 \\
57359.3 & 6 & 11.10 & 11.07 & 11.13 & 4.25 & 4.22 & 4.28 & 15.42 & 15.37 & 15.48 \\
57362.2 & 6 & 11.12 & 11.08 & 11.16 & 4.28 & 4.25 & 4.32 & 15.37 & 15.31 & 15.43 \\
57368.9 & 6 & 11.06 & 11.02 & 11.09 & 4.24 & 4.21 & 4.27 & 15.42 & 15.36 & 15.48 \\
57371.4 & 6 & 11.03 & 10.99 & 11.06 & 4.25 & 4.22 & 4.29 & 15.38 & 15.32 & 15.44 \\
57374.4 & 6 & 11.03 & 11.00 & 11.07 & 4.25 & 4.22 & 4.28 & 15.39 & 15.34 & 15.45 \\
57377.2 & 6 & 11.01 & 10.97 & 11.04 & 4.25 & 4.22 & 4.28 & 15.38 & 15.32 & 15.44 \\
57380.9 & 6 & 11.03 & 10.99 & 11.06 & 4.26 & 4.23 & 4.29 & 15.37 & 15.31 & 15.43 \\
57383.2 & 6 & 10.97 & 10.94 & 11.00 & 4.22 & 4.19 & 4.25 & 15.41 & 15.35 & 15.47 \\
57385.9 & 6 & 11.02 & 10.98 & 11.05 & 4.25 & 4.22 & 4.28 & 15.38 & 15.32 & 15.43 \\
\hline
\end{tabular}
\label{tab:table5}
\end{minipage}
\end{table*}

\begin{table*}
\begin{minipage}{\textwidth}
\centering
\renewcommand{\arraystretch}{1.2}
\begin{tabular}{ccccccccccc}
\hline
MJD & $N_{\text{obs}}$ & $\log(L)$ & $\log(L_{-})$  &  $\log(L_{+})$ & $\log(T)$  & $\log(T_{-})$  & $\log(T_{+})$ & $\log(R)$  & $\log(R_{-})$  & $\log(R_{+})$ \\
\hline
57393.3 & 6 & 10.99 & 10.96 & 11.02 & 4.24 & 4.21 & 4.27 & 15.38 & 15.33 & 15.44 \\
57396.7 & 6 & 10.93 & 10.89 & 10.96 & 4.22 & 4.19 & 4.25 & 15.39 & 15.33 & 15.45 \\
57399.2 & 6 & 10.94 & 10.91 & 10.97 & 4.22 & 4.19 & 4.25 & 15.40 & 15.35 & 15.46 \\
57427.7 & 6 & 10.80 & 10.77 & 10.83 & 4.21 & 4.18 & 4.24 & 15.35 & 15.29 & 15.40 \\
57429.4 & 6 & 10.82 & 10.79 & 10.85 & 4.20 & 4.17 & 4.23 & 15.38 & 15.32 & 15.44 \\
57433.1 & 6 & 10.75 & 10.72 & 10.79 & 4.16 & 4.14 & 4.19 & 15.42 & 15.37 & 15.48 \\
57437.1 & 6 & 10.80 & 10.77 & 10.83 & 4.19 & 4.17 & 4.22 & 15.39 & 15.33 & 15.44 \\
57447.2 & 6 & 10.77 & 10.73 & 10.80 & 4.19 & 4.16 & 4.22 & 15.37 & 15.32 & 15.43 \\
57450.0 & 6 & 10.73 & 10.70 & 10.76 & 4.17 & 4.15 & 4.20 & 15.39 & 15.34 & 15.45 \\
57453.9 & 6 & 10.79 & 10.76 & 10.82 & 4.21 & 4.18 & 4.24 & 15.35 & 15.29 & 15.40 \\
57456.0 & 6 & 10.74 & 10.71 & 10.77 & 4.22 & 4.20 & 4.25 & 15.29 & 15.24 & 15.35 \\
57457.5 & 5 & 10.69 & 10.66 & 10.72 & 4.16 & 4.13 & 4.19 & 15.39 & 15.34 & 15.45 \\
57465.1 & 6 & 10.71 & 10.67 & 10.74 & 4.18 & 4.16 & 4.21 & 15.36 & 15.30 & 15.41 \\
57470.2 & 6 & 10.74 & 10.71 & 10.77 & 4.20 & 4.17 & 4.23 & 15.34 & 15.28 & 15.39 \\
57472.2 & 6 & 10.72 & 10.69 & 10.75 & 4.18 & 4.16 & 4.21 & 15.37 & 15.31 & 15.42 \\
57475.6 & 6 & 10.72 & 10.68 & 10.75 & 4.21 & 4.18 & 4.23 & 15.32 & 15.26 & 15.37 \\
57477.2 & 6 & 10.71 & 10.68 & 10.74 & 4.18 & 4.15 & 4.20 & 15.37 & 15.32 & 15.43 \\
57480.0 & 6 & 10.64 & 10.61 & 10.67 & 4.17 & 4.15 & 4.20 & 15.34 & 15.29 & 15.40 \\
57487.1 & 6 & 10.65 & 10.62 & 10.68 & 4.18 & 4.15 & 4.20 & 15.34 & 15.29 & 15.40 \\
57489.0 & 6 & 10.63 & 10.60 & 10.66 & 4.18 & 4.16 & 4.21 & 15.32 & 15.26 & 15.37 \\
57499.7 & 4 & 10.56 & 10.52 & 10.60 & 4.15 & 4.12 & 4.18 & 15.34 & 15.28 & 15.41 \\
57502.7 & 6 & 10.63 & 10.60 & 10.66 & 4.19 & 4.17 & 4.22 & 15.30 & 15.25 & 15.36 \\
57505.2 & 6 & 10.63 & 10.60 & 10.66 & 4.18 & 4.16 & 4.21 & 15.31 & 15.26 & 15.37 \\
57508.9 & 6 & 10.65 & 10.62 & 10.68 & 4.19 & 4.16 & 4.22 & 15.32 & 15.26 & 15.37 \\
57511.8 & 6 & 10.59 & 10.55 & 10.62 & 4.18 & 4.16 & 4.21 & 15.30 & 15.25 & 15.36 \\
57516.1 & 6 & 10.61 & 10.58 & 10.64 & 4.19 & 4.16 & 4.21 & 15.30 & 15.24 & 15.36 \\
57520.3 & 6 & 10.55 & 10.52 & 10.58 & 4.18 & 4.15 & 4.21 & 15.29 & 15.23 & 15.34 \\
57523.7 & 6 & 10.57 & 10.54 & 10.60 & 4.19 & 4.17 & 4.22 & 15.27 & 15.21 & 15.32 \\
57528.4 & 6 & 10.55 & 10.52 & 10.58 & 4.19 & 4.16 & 4.21 & 15.27 & 15.21 & 15.33 \\
57533.6 & 6 & 10.51 & 10.48 & 10.55 & 4.13 & 4.11 & 4.15 & 15.37 & 15.31 & 15.42 \\
57552.5 & 6 & 10.48 & 10.45 & 10.51 & 4.19 & 4.17 & 4.22 & 15.22 & 15.17 & 15.28 \\
57554.8 & 6 & 10.62 & 10.59 & 10.66 & 4.10 & 4.08 & 4.12 & 15.48 & 15.43 & 15.53 \\
57557.2 & 6 & 10.43 & 10.40 & 10.46 & 4.19 & 4.17 & 4.22 & 15.20 & 15.14 & 15.26 \\
57560.0 & 6 & 10.47 & 10.44 & 10.50 & 4.23 & 4.21 & 4.26 & 15.14 & 15.08 & 15.19 \\
57563.0 & 6 & 10.51 & 10.47 & 10.54 & 4.20 & 4.17 & 4.23 & 15.22 & 15.16 & 15.28 \\
57569.1 & 6 & 10.61 & 10.57 & 10.64 & 4.09 & 4.07 & 4.11 & 15.50 & 15.44 & 15.55 \\
57577.8 & 6 & 10.48 & 10.45 & 10.51 & 4.20 & 4.17 & 4.23 & 15.21 & 15.15 & 15.26 \\
57591.7 & 6 & 10.35 & 10.32 & 10.38 & 4.13 & 4.11 & 4.15 & 15.29 & 15.23 & 15.34 \\
57600.9 & 6 & 10.34 & 10.31 & 10.38 & 4.15 & 4.13 & 4.17 & 15.24 & 15.19 & 15.30 \\
57608.5 & 6 & 10.32 & 10.29 & 10.36 & 4.20 & 4.18 & 4.23 & 15.13 & 15.07 & 15.18 \\
57611.5 & 6 & 10.25 & 10.22 & 10.28 & 4.18 & 4.16 & 4.21 & 15.13 & 15.07 & 15.18 \\
57615.4 & 6 & 10.32 & 10.29 & 10.35 & 4.23 & 4.21 & 4.26 & 15.06 & 15.00 & 15.12 \\
57622.5 & 5 & 10.29 & 10.26 & 10.33 & 4.21 & 4.18 & 4.24 & 15.09 & 15.04 & 15.15 \\
57626.7 & 5 & 10.19 & 10.15 & 10.22 & 4.17 & 4.14 & 4.19 & 15.13 & 15.07 & 15.19 \\
57630.4 & 4 & 10.33 & 10.29 & 10.38 & 4.24 & 4.21 & 4.27 & 15.06 & 15.00 & 15.12 \\
57632.8 & 4 & 10.32 & 10.27 & 10.37 & 4.26 & 4.23 & 4.30 & 15.00 & 14.94 & 15.06 \\
57640.0 & 6 & 10.21 & 10.18 & 10.24 & 4.18 & 4.15 & 4.21 & 15.12 & 15.06 & 15.17 \\
57645.5 & 6 & 10.18 & 10.15 & 10.21 & 4.19 & 4.16 & 4.21 & 15.09 & 15.03 & 15.14 \\
57654.2 & 5 & 10.27 & 10.24 & 10.31 & 4.19 & 4.16 & 4.22 & 15.12 & 15.06 & 15.18 \\
57667.3 & 3 & 10.21 & 10.16 & 10.25 & 4.19 & 4.16 & 4.22 & 15.09 & 15.04 & 15.15 \\
57677.3 & 3 & 10.20 & 10.15 & 10.24 & 4.18 & 4.15 & 4.21 & 15.11 & 15.05 & 15.17 \\
57680.2 & 4 & 10.27 & 10.23 & 10.31 & 4.21 & 4.18 & 4.24 & 15.09 & 15.03 & 15.15 \\
57682.5 & 6 & 10.18 & 10.15 & 10.21 & 4.17 & 4.15 & 4.20 & 15.12 & 15.06 & 15.17 \\
57687.0 & 3 & 10.21 & 10.17 & 10.26 & 4.22 & 4.19 & 4.25 & 15.04 & 14.98 & 15.09 \\
57688.9 & 5 & 10.21 & 10.18 & 10.25 & 4.23 & 4.20 & 4.26 & 15.01 & 14.96 & 15.07 \\
57704.0 & 2 & 10.27 & 10.19 & 10.36 & 4.05 & 4.01 & 4.08 & 15.41 & 15.31 & 15.51 \\
\hline
\end{tabular}
\end{minipage}
\end{table*}

\end{document}